\documentclass{aa}
\usepackage{txfonts}
\usepackage[authoryear]{natbib}
\usepackage{graphicx}
\graphicspath{{converted_graphics/}}

\def \s {Cyg~X-3$\;$}

\def \cyg{Cyg~X-3 }
\def \cygp{Cyg~X-3}

\def \aa {A$\&$A$\;$}

\def \degmark{^\circ}

\def \phcmsec{\hbox{ph. cm$^{-2}$ s$^{-1}$}}
\def \ctscmsec{\hbox{cts. cm$^{-2}$ s$^{-1}$}}

\def \gray {$\gamma$-ray }
\def \enmev {$E > 100 \: MeV$}

\def\fermi {{\it Fermi}}
\def\fermip {{\it Fermi }}
\def\swift {{\it Swift }}
\def\mt {   }

\def\aba { }
\def\abm { }
\def\abc  { }

\def\abe  {   }

\def\abh  { }

\begin{document}

\title{AGILE detection of Cygnus X-3 $\gamma$-ray active states during the period  mid-2009/mid-2010}

\date{}          % Enter your date or \today between curly braces
\authorrunning {A.Bulgarelli et al.}
\titlerunning { AGILE detection of Cygnus X-3 during the period mid-2009/mid-2010}

\author{A.~Bulgarelli\inst{1},
  M.~Tavani\inst{2,3,5,10},  A.W.~Chen\inst{3,4}, Y.~Evangelista\inst{2}, M.~Trifoglio\inst{1}, F.~Gianotti\inst{1},
  G.~Piano\inst{2,10}, S. Sabatini\inst{5, 10}, E. Striani\inst{5, 10},
  G. ~Pooley\inst{18}, S.~Trushkin\inst{19}, N.~A.~Nizhelskij\inst{19}, M.~McCollough\inst{20}, K.~I.~I.~Koljonen\inst{22},
  D.~Hannikainen\inst{21,22,26},   A.~L\"{a}hteenm\"{a}ki \inst{22}, J.~Tammi\inst{22}, N.~Lavonen\inst{22}, D.~Steeghs\inst{24},
  A. Aboudan\inst{25}, A.~Argan\inst{2}, G.~Barbiellini\inst{6}, R.~Campana\inst{2}
  P.~Caraveo\inst{4}, P.~W.~Cattaneo \inst{9},
  V.~Cocco\inst{2},
  T. Contessi\inst{3,4},
  E.~Costa\inst{1}, F.~D'Ammando\inst{23},
  E.~Del Monte\inst{2}, G.~De Paris\inst{2},
  G.~Di~Cocco\inst{1}, I.~Donnarumma\inst{2},
  M.~Feroci\inst{2}, M.~Fiorini\inst{4},
  F.~Fuschino\inst{3}, M.~Galli\inst{7},
  A.~Giuliani\inst{4}, M.~Giusti\inst{2,3}, C.~Labanti\inst{1}, I.~Lapshov\inst{2}, F.~Lazzarotto\inst{2},
  P.~Lipari\inst{8, 16}, F.~Longo\inst{6},
  M.~Marisaldi\inst{1}, S.~Mereghetti\inst{4}, E.~Morelli\inst{1}, E. Moretti\inst{6},
  A.~Morselli\inst{10}, L.~Pacciani\inst{2},
  A.~Pellizzoni\inst{17}, F.~Perotti\inst{4},
  P.~Picozza\inst{5,10}, M. Pilia\inst{11,17}, M.~Prest\inst{11},
  G.~Pucella\inst{12}, M.~Rapisarda\inst{12}, A.~Rappoldi \inst{9}, A. Rubini\inst{2}
  P.~Soffitta\inst{2},  A.~Trois\inst{2},
  E.~Vallazza\inst{6}, S.~Vercellone\inst{23}, V.~Vittorini\inst{2,3},
  A.~Zambra\inst{4}, D.~Zanello\inst{8},
  P.~Giommi\inst{13},
  C.~Pittori\inst{13}, F.~Verrecchia\inst{13},
  P.~Santolamazza\inst{13}, F.~Lucarelli\inst{13},
  S.~Colafrancesco\inst{13, 27, 28}, L.~Salotti\inst{11}
}

\institute{$^1$ INAF/IASF--Bologna, Via Gobetti 101, I-40129 Bologna, Italy \\
$^2$ INAF/IASF--Roma, Via del Fosso del Cavaliere 100, I-00133 Roma, Italy \\
$^3$ CIFS--Torino, Viale Settimio Severo 3, I-10133, Torino, Italy \\
$^4$ INAF/IASF--Milano, Via E.~Bassini 15, I-20133 Milano, Italy \\
$^5$ Dip. di Fisica, Univ. ``Tor Vergata'', Via della Ricerca Scientifica 1, I-00133 Roma, Italy \\
$^6$ Dip. di Fisica and INFN Trieste, Via Valerio 2, I-34127 Trieste, Italy\\
$^7$ ENEA--Bologna, Via Biancafarina 2521, I-40059 Medicica (BO), Italy\\
$^8$ INFN--Roma ``La Sapienza'', Piazzale A. Moro 2, I-00185 Roma, Italy\\
$^9$ INFN--Pavia, Via Bassi 6, I-27100 Pavia, Italy\\
$^{10}$ INFN--Roma ``Tor Vergata'', Via della Ricerca Scientifica 1, I-00133 Roma, Italy\\
$^{11}$ Dip. di Fisica, Univ. dell'Insubria, Via Valleggio 11, I-22100 Como, Italy\\
$^{12}$ ENEA--Frascati, Via E. Fermi 45, I-00044 Frascati (Roma), Italy\\
$^{13}$ ASI--ASDC, Via G. Galilei, I-00044 Frascati (Roma), Italy\\
$^{14}$ ASI, Viale Liegi 26 , I-00198 Roma, Italy\\
$^{15}$ Dipartimento di Fisica Nucleare e Teorica, Universit\'a di Pavia, Via Bassi 6, Pavia, I-27100, Italy\\
$^{16}$ Dipartimento di Fisica, Universita\'a La Sapienza, Piazza Aldo Moro 2, I-00185 Roma, Italy\\
$^{17}$ INAF Osservatorio Astronomico di Cagliari, Poggio dei Pini, I-09012 Capoterra (Cagliari), Italy\\
$^{18}$ Astrophysics Group, Cavendish Laboratory, 19 J. J. Thomson Avenue, Cambridge CB3 0HE, UK\\
$^{19}$ Special Astrophysical Observatory RAS, Karachaevo-Cherkassian Republic, Nizhnij Arkhyz 36916, Russia\\
$^{20}$ Smithsonian Center for Astrophysics, 60 Garden Street, Cambridge, Massachusetts 02138, USA\\
$^{21}$ Finnish Centre for Astronomy with ESO (FINCA), University of Turku, V\"{a}is\"{a} l\"{a} ntie 20, FI-21500 Piikki\"{o}, Finland\\
    $^{22}$Aalto University Mets\"ahovi Radio Observatory Mets\"ahovintie 114 FIN-02540 Kylm\"al\"a Finland\\
$^{23}$ INAF/IASF--Palermo, Via Ugo La Malfa 153, I-90146 Palermo, Italy\\
$^{24}$ Astronomy And Astrophysics Department of Physics, University of Warwick, Coventry CV4 7AL, UK\\
$^{25}$ CISAS ``G. Colombo'' , Univ. di Padova, Via Venezia 15, 35131 Padova, Italy  \\
$^{26}$ Department of Physics and Space Sciences, Florida Institute of Technology, 150 W. University Blvd., Melbourne, FL 32901, USA \\
$^{27}$ INAF - Osservatorio Astronomico di Roma, Via Frascati 33, I-00040 Monteporzio, Italy\\
$^{28}$ School of Physics, University of the Witwatersrand, Johannesburg Wits 2050, South Africa\\
}

\offprints{A. Bulgarelli, \email{bulgarelli@iasfbo.inaf.it} }
\date{accepted}

\abstract{Cygnus X-3 (\cygp) is a  well-known  microquasar producing
variable emission at all wavelengths. \cyg  is a
prominent X-ray binary producing relativistic jets, and studying
its high energy emission is crucial for the understanding of the
fundamental acceleration processes in accreting compact objects.}
{Our goal is to study extreme particle acceleration and $\gamma$-ray
production above 100 MeV during special spectral states of \cyg
usually characterized by a low hard X-ray flux and enhanced
soft X-ray states. } 
{We observed \cyg with the AGILE satellite
in extended time intervals from \abe{2009 Jun.--Jul, and 2009 Nov.--2010 Jul}. 
We report here the results of the AGILE $\gamma$-ray
monitoring of \cyg
 as well as the results from  extensive
multiwavelength campaigns involving radio (RATAN-600, AMI-LA and Mets\"{a}hovi Radio Observatories)
and X-ray monitoring data (XTE and Swift). }
{We detect a series
of repeated $\gamma$-ray flaring activity from \cyg that
correlate
with the  soft X-ray states and episodes of decreasing or
non-detectable hard X-ray emission. Furthermore, we detect
$\gamma$-ray enhanced emission that tends to be associated with radio
flares greater than 1 Jy at 15 GHz, confirming a trend already
detected in previous observations. The source remained active
above 100 MeV for an extended period of time (almost 1.5
months in 2009 Jun.--Jul and 1 month in 2010 May). We study in
detail the short timescale $\gamma$-ray flares that occurred before
or near the radio peaks.} {Our results confirm the transient
nature of the extreme particle acceleration from the microquasar
\cygp. A series of repeated $\gamma$-ray flares shows
correlations with radio and X-ray
emission confirming a well established trend of emission.
We compare our results with Fermi-LAT
and MAGIC TeV
observations of \cygp.}


\maketitle

\section{Introduction}

\object{Cygnus X-3} (Cyg X-3) is one of the most puzzling and interesting
compact objects in our Galaxy. Discovered in 1966 \cite{giacconi},
\cyg is a high-mass X-ray binary exhibiting a 4.8 hr
modulation in its X-ray  \cite{parsignault, vilhu}, infrared
\cite{mason,vankerkwijk,vankerkwijk2} and $\gamma$-ray
\cite{Abdo_2009b}  emission. The system is believed to be
composed of a mass-donating Wolf-Rayet star
\cite{vankerkwijk2} orbiting around a compact object (neutron
star or black hole). Certainly what makes \cyg an
\abe{interesting} object is its ability to produce very energetic
relativistic radio jets \cite{molnar,miodu} that usually occur in
coincidence with particular spectral  states
\cite{mccollough,zdziarski,szostek,hjalmarsdotter,koljonen}.
Many decades of X-ray monitoring of \cyg have provided a wealth of
information on this source, and in general the soft, intermediate
and hard X-ray states show   correlated or anti-correlated
behaviours with respect  to the radio and jet emission. In particular,
the soft (1-10 keV) and hard (20-100 keV) X-ray emission
from \cyg are clearly anti-correlated during ``normal'' stages
during which the source produces radio emission at  low or
intermediate levels with no major radio jet production.
Major radio flares in \cyg are preceded by quenched states, during
which the source is in a soft X-ray state  and radio emission
is strongly suppressed.

The ability of \cyg to efficiently accelerate particles in relativistic jets,
and the favourable jet geometry
parameters that make the transient jet phenomenon very dramatic  (e.g.
\cite{miodu} )  have attracted
considerable attention from the astrophysics community. \cyg is a  Galactic ``microquasar'',
and the inner dynamics and accretion properties of its compact object
can shed light on fundamental physical processes of relativistic jet sources, both Galactic
and extragalactic. From this point of view, exploring the high-energy
emission from \cyg and unveiling its temporal and physical properties open the
way to a detailed understanding of the plasma properties of inner accretion disks
of compact objects and of particle acceleration processes under extreme radiative conditions.

Motivated by these reasons, our group has made a detailed study of
$\gamma$-ray emission from \cyg with the AGILE satellite. The AGILE discovery of transient
$\gamma$-ray emission from \cyg in 2008 Apr. associated with a specific spectral state preceding
a major radio jet ejection opened a new window of investigation of microquasars.
Several other major $\gamma$-ray emission episodes from \cyg
have been detected by AGILE and Fermi since 2008 \cite{Tavani_2009b, Abdo_2009b}.
In this paper, we focus on the main results of our extensive search for transient
$\gamma$-ray emission from \s carried out in the energy range 100  MeV - 50 GeV by  AGILE
during the periods 2009 Jun--Jul and 2009 Dec.--2010 mid-Jun.
We find that the activity in $\gamma$-rays during the
periods 2009 Jun--Jul and 2010 May  temporally coincide with the hard X-ray emission reach minimum values. 
Our results confirm the
overall trend found by \citep{Tavani_2009b}, i.e., that the  highest
$\gamma$-ray emission above 100 MeV from \cyg  is associated with soft X-ray
spectral states
that are in general coincident with or anticipate radio jet ejections. Furthermore, our
results show that $\gamma$-ray emission from \cyg is detectable by AGILE not only during
relatively short (1-2 day) flares as in \citep{Tavani_2009b} but also during ``extended''
periods lasting
several days or weeks (as during 2009 Jun-Jul). Detecting continuous
$\gamma$-ray emission during ``active'' phases is  of great theoretical relevance for the
 modelling of \cyg.

We also briefly compare here the AGILE and \fermi-LAT results. We
also consider a recent investigation by the MAGIC group reporting
their results on a monitoring program of \cyg  in the TeV energy range
\cite{magic}, and briefly discuss the implications with respect
to the AGILE detections.

\section{The AGILE GRID observations of \s}

AGILE (Astrorivelatore Gamma ad Immagini LEggero - Light Imager for Gamma-ray Astrophysics)
is a scientific mission of the Italian Space Agency (ASI)  launched on April 23, 2007 \cite{Tavani_2009a}.
The AGILE scientific payload is made of three  detectors: (1)  a $\gamma$-ray imager
made of a Tungsten-Silicon Tracker (ST) \cite{Barbiellini_2002, Prest_2003,Bulgarelli_2010} with a large
field of view (about 60$^{\circ}$);
(2) a co-axial  hard X-ray silicon detector (named
Super-AGILE  \cite{Feroci2007}), for imaging in the 18-60~keV
energy range,  and (3) a CsI(Tl) Mini-Calorimeter (MCAL) detector
\cite{Labanti_2006} that detects $\gamma$-rays or particle energy
deposits between $\sim$350~keV and 100~MeV. The whole instrument is surrounded
by an anti-coincidence (AC) system \cite{Perotti_2006} of plastic
scintillators for the rejection of background charged particles. An effective background
rejection, event trigger logic, and on-board data storage and transmission is implemented
\cite{argan}.
ST, MCAL and AC form the
so called Gamma-Ray Imaging Detector (GRID) for observations in
the 30~MeV-50~GeV $\gamma$ energy range. The AGILE orbital characteristics
(quasi-equatorial with an inclination angle of 2.5 degrees and average 530 km altitude)
are optimal for low-background $\gamma$-ray observations. AGILE data are transmitted
to the ASI Malindi ground station in Kenya, and quickly transferred to the ASI Science Data Center
(ASDC) near Frascati. Data processing of $\gamma$-ray data is then
carried out at the ASDC and AGILE Team locations.

The AGILE-GRID is optimized in the 100 MeV - 1 GeV range, as
demonstrated by the calibrated  data \citep{Cattaneo_2011, Chen_2011b} and by the in-orbit
performance \cite{Tavani_2009a}.

The AGILE $\gamma$-ray exposure of the Cygnus region can be divided in two parts.

(1) A first set of exposures obtained in the satellite ``pointing mode'' characterized
 by pointing centroids near the Cygnus region in the Galactic plane.
 In this mode, AGILE accumulated a total effective time   of
 $\sim$338 days during the period 2007 Jul--2009 Oct. Table~\ref{table_1}
 provides the details of the main time intervals analyzed in this paper for which a substantial exposure was
 accumulated in the Cygnus region in this pointing mode.

 (2) A second set of exposures
 obtained quasi-continuously with the satellite operating in  ``spinning mode'' since 
 2009 Nov. In this mode, the satellite axis sweeps an entire circle in the sky in approximately
 7 minutes. Depending on the season, the whole sky is progressively exposed with a typical
 accumulating pattern. The Cygnus region is favorably positioned in the sky, and has been
 continuously monitored since   2009 Nov. with a 2-days  exposure comparable with
 1-day pointing exposure level.

\begin {table*}[!htb]
\caption {\em{AGILE observations of the \s region analyzed in this paper.}}
\label{table_1}
\renewcommand{\arraystretch}{1.2} % enlarge line spacing
\begin{tabular}{@{}llllllllcclll}
\hline
\textbf {Obs. block} & \textbf{ l } & \textbf{ b } & \textbf{ Time (MJD) } &
\textbf { Mean off-axis angle }  & $\sqrt{T_s}$ & Flux  \\
\hline

OB7500/OB7600  & 92.835  &  -9.574 & 54997.50-55027.50  &  17/29 & 3.5  & 32 $\pm$ 10  \\

OB7700  & 105.729 &   7.231 & 55027.50-55055.50  &  30  &  $<$2  &  $<$ 20  \\

spinning & - & - & 55168.50-55362.50 & - &  $<$2 & $<$ 16 & \\

\hline
\end{tabular}

\vskip .1in
 The Table provides: (1) the observation block (OB)
number; (2-3) the galactic coordinates $l$ and $b$ of the pointing
centroids; (4) the time interval in MJD; (5) the off-axis position
of \cyg at
the beginning of the OB \aba{(in degrees)}; (6) the statistical significance $\sqrt{T_s}$ of the {source detection according to the maximum } likelihood ratio test; (7) the
 period-averaged flux $F$ (\enmev) in $10^{-8}$ \phcmsec (if $\sqrt{T_s}<3$ a $2-\sigma$ upper limit
is reported)
\end{table*}

\subsection{Data analysis method}
\label{sec::analysis}
The data analysis was performed on the data set in pointing mode
generated with the reprocessing no. 3 of the AGILE Standard
Pipeline, and with the AGILE-GRID software package version 4
publicly available at the ASI Data Center web site
(http://agile.asdc.asi.it/). The analysis has been performed with
the FM3.119 filter and the calibration matrix used is the I0023
version. The events collected during the passage in the
South-Atlantic Anomaly and the Earth albedo background were
consistently rejected. The GRID event directions were reconstructed
by  the AGILE Kalman filter technique \citep{Giuliani_2006}. To
reduce the particle background contamination, we selected only
events flagged as confirmed $\gamma$-ray events (\textit{G} class
events, corresponding to an on-axis effective area of $\sim 350 \:$
cm$^2$  at 100 MeV). 

The multi-source likelihood analysis method (MSLA) \aba {\citep{Mattox_1996a}} was used to search
for persistent and transient emission from Cygnus region; this analysis method iteratively
optimizes position, flux and spectral index of all the sources of the region.
The likelihood ratio $T_s$ is then simply the ratio of  the maximum likelihood of two hypothesis (e.g. the absence and the presence of a source).
For this analysis, integrating all flaring episodes in 2009 Jun-Jul, we obtained 
(see Section \ref{sec:spectra} and Fig. \ref{fig-5}) and kept fixed the
photon  index of 2.0 for \cygp. This is consistent with other AGILE detections
reported in \citep{Tavani_2009b}.During the analysis the position of the source is kept
free and constrained with the 95$\%$ confidence contour level. The AGILE photon counts, exposure, and Galactic background maps
were generated with a bin size of $0.3^{\circ} \times 0.3^{\circ}$
for E$>$100 MeV to compute the period-averaged source flux and its
evolution. The analysis was performed over a region of
$10\degmark$ radius.

The Galactic diffuse $\gamma$-ray radiation \citep{Giuliani_2004}
and the isotropic emission are taken into account in the model.
The Galactic diffuse emission model is based on a 3D grid
with bin of 0.25$\degmark$ in galactic longitude and latitude and
0.2 kpc in distance along the line of sight. In order to model the
matter distribution in the galaxy we use the HI
\citep{2005A&A...440..775K}  and CO \citep{Dame_2001} radio survey
and  we take into account the cosmic ray models
\citep{Chi_1991} which can differ from the locally observed cosmic
ray spectrum.

Two energy bands have been considered in the analysis:
100 MeV - 50 GeV and 400 MeV - 50 GeV.

 Only the detections with  $\sqrt{T_s} \ge 3$ and with a position
 consistent with  \s source are selected.  For this subclass of  selected flare peaks,
a non-automatic verification is performed to further confirm the results.
Pre-trial significance was determined by calculating 
the $T_s$ density function of our analysis method \citep{Mattox_1996a} by means of Monte Carlo simulation \citep{Bulgarelli_2011}.
For the determination of the post-trial significance we have taken  into account
only the number of  bins of the light curve\footnote{This means that we
have considered $58$ maps in pointing mode and $92$ maps in spinning mode.
The total number of maps is 150.}.

\begin {table*}[!htb]
\caption {\em{List of Cygnus region sources with E $>$ 100 MeV.}}
\label{table_2}
\renewcommand{\arraystretch}{1.2}
\begin{tabular}{@{}llllllllll} \hline
&  &   & E$>$100 MeV &  & E$>$400 MeV & &  \\
\textbf{ AGILE Name } & \textbf{ l } & \textbf{ b }  & \textbf{ $\sqrt{T_s}$ } & Flux  & \textbf{ $\sqrt{T_s}$ }
& Flux & \textbf{Counterpart name} \\
AGL 2021+4029 & 78.24 & 2.16  & 42.1 & 141  $\pm$ 4  & 30  & 33 $\pm$ 1.5  & \object{1AGL J2022+4032} - Gamma Cygni \\
AGL 2021+3652 & 75.24 & 0.14  & 23.3 & 67  $\pm$ 3  & 21 &  19 $\pm$ 1.2 & \object{1AGL J2021+3652} - PSR J2021+3651 \\
AGL 2030+4129 & 80.11 & 1.25  & 8.1 & 18  $\pm$ 4 & 8.7 & 5.6 $\pm$ 0.7  & \object{1AGL J2032+4102} - LAT PSR J2032+4127 \\
AGL 2026+3346 & 73.28 & -2.49  & 6.8 & 10 $\pm$ 1.7 & 6.2 & 2.8 $\pm$ 0.5 & - \\
AGL 2046+5032 & 88.99 & 4.54  & 6.5 & 10 $\pm$ 1.7 & 6.1 & 2.6 $\pm$ 0.5 & - \\
AGL 2016+3644 & 74.59 & 0.83  & 6.3 & 14 $\pm$ 2.3 & - & - & - \\
AGL 2029+4403 & 81.97 & 3.04  & 5.4 & 14 $\pm$ 3 & 5.1 & 3.4 $\pm$ 0.7 & -\\
AGL 2033+4056 & 79.92 & 0.58  & 5.3 & 15 $\pm$ 2 & 3.1 & 2.3 $\pm$ 0.7 & \object{1AGL J2032+4102} - Cygnus X-3 \\
AGL 2038+4313 & 82.32 & 1.18  & 5.1 & 15 $\pm$ 3 & - & - & - \\
AGL 2024+4027 & 78.56 & 1.63  & 5.0 & 24 $\pm$ 5 & 6.1 & 11 $\pm$ 2 & - \\
AGL 2019+3816 & 76.24 & 1.14  & 4.2& 11 $\pm$ 2.4 & 4.2 & 2 $\pm$ 0.6 & - \\
AGL 2036+3954 & 79.47 & -0.56  & 3.4 & 5.0 $\pm$ 1.5 & 5.7 & 3.8 $\pm$ 0.7 & - \\
\hline
\end{tabular}

\vskip .1in
 The Table provides: (1) AGILE name of the sources; (2) (3) the galactic coordinates $l$ and $b$; (4) (6)
 the statistical significance $\sqrt{T_s}$ of the {source detection according to the maximum }
 likelihood ratio test for E$>$100 MeV and E$>$400 MeV respectively; (5) (7) the
 period-averaged flux $F$ (\enmev) in $10^{-8}$ \phcmsec for E$>$100 MeV and E$>$400 MeV respectively;
 (8) a possible counterpart

\end{table*}

The Cygnus region is characterized by  complex $\gamma$-ray emission: several
 $\gamma$-ray sources are detected above 100 MeV, and it is important to
correctly model  the diffuse radiation of the region.
Three  bright  $\gamma$-ray sources dominate the Cygnus region and also reported in the First AGILE Catalog  \cite{Pittori_2009}.
 They are all  $\gamma$-ray pulsars \cite{halpern,Abdo_2009a,camilo}.
We added more nearby sources in the $\gamma$-ray model (Table \ref{table_2}) and the statistical analysis of these sources
is compared  with the First AGILE Catalog analysis. A  new likelihood analysis investigation
 has been performed to take into account a new background event filter (FM3.119).
 We have also added  considerably more exposure to the data set.

For our MSLA method we first determined self-consistently the positions of all bright sources
(including also a \cyg candidate source,  AGL 2033+4056) by considering the whole
set of AGILE-GRID data in pointing mode for the Cygnus region. Subsequently (see
next Section), for the time-resolved analysis of  \cyg candidate source, we
 fixed the fluxes and the positions of the
brightest sources as shown in Table \ref{table_2}, keeping the position and flux of \cyg free.
In particular, we studied  the $\gamma$-ray source
\footnote{Notice that the
initially reported AGILE First Catalog source 1AGL~J2032+4102 \cite{Pittori_2009}
comprises two $\gamma$-ray sources separated by only 0.4$\degmark$
as clarified in \citep{Tavani_2009b} and confirmed by our more refined
analysis. }
that we call here  AGL 2030+4129 that is
positioned only  0.4$\degmark$ from  \s position. This $\gamma$-ray source is
consistent in position and average flux
with the source \object{1FGL~J2032.2+4127} \cite{1FGL}
which is the counterpart of  the radio pulsar \object{PSR~J2032+4127}
(of spin period $P = 147$~ms, and a most likely distance $d \simeq 1.8$~kpc
\citep{camilo}).
Given the differences in the published determinations\footnote{
\fermi-LAT First Catalog analyses give a $\gamma$-ray photon spectral index consistent
with the value $\alpha = 2.24 \pm 0.034$. However,
an analysis of the $\gamma$-ray \fermi-LAT data of PSR J2032+4127
that was carried out by selecting only
the on-pulse phase photons gives an exponentially cut-off power law spectrum of
 photon spectral index $\alpha = 1.1 \pm 0.2 \pm 0.2$ and cutoff energy $E_c = (3.0 \pm 0.6 \pm 0.7) \,$GeV for a pulsed $\gamma$-ray flux of $F_g = (7 \pm 1 \pm 1)\times 10^{-8} \, \rm cm^{-2} \,
 s^{-1}$ above 100 MeV \citep{camilo}.} of the $\gamma$-ray spectral index of
 AGL 2030+4129/1FGL~J2032.2+4127, we
considered in our analysis two $\gamma$-ray spectral indices
$\alpha_1 =  2.24$ (\fermi-LAT First Catalog) , and $\alpha_2 =
1.84  \pm 0.2$ calculated with AGILE data; for these indices we
find the corresponding average fluxes $26  \pm 3$ and  $18 \pm 4
\cdot 10^{-8}$ \phcmsec. In this paper we adopt the spectral index
value $\alpha = 1.84 \pm 0.2$ \abe{that is the value calculated with the AGILE data}.

An alternative method  using the False Discovery Rate (FDR, \citep{ Benjamini_1995, Miller_2001, Hopkins_2002}) has been used to analyze the flares detected with MSLA. The detection method (FDRM) is a statistical test
that  takes into account the corrections for multiple testing, as
needed for example in repeated systematic searches. The FDRM allows to control the expected rate of false detections (due to background fluctuations) within a selected sample. The FDR-$\alpha$ parameter provides the fraction of expected false detections for a given source selection. The method was adapted to the analysis of AGILE gamma-ray data of the Galactic plane \cite{Sabatini_2010, Tavani_2009b} in pointing mode and for E$>$100 MeV (for spinning mode end for E$>$400 MeV the method is not available). The FDRM is complementary and more conservative than the determination of a post-trial detection significance based on a simulated set of replicated photon maps;  applying  FDR method to typical short-time  AGILE-GRID photon maps in the Galactic plane, FDR-$\alpha$ values near or below 0.01 fully qualify for gamma-ray transients (it correspond to post-trial random occurrences of 1-day map replications equal to 1 every 300 or more). 

\subsection{Flaring $\gamma$-ray activity observed by AGILE}

\begin{figure*}[!htb]
\centering
 \includegraphics[width=16cm]{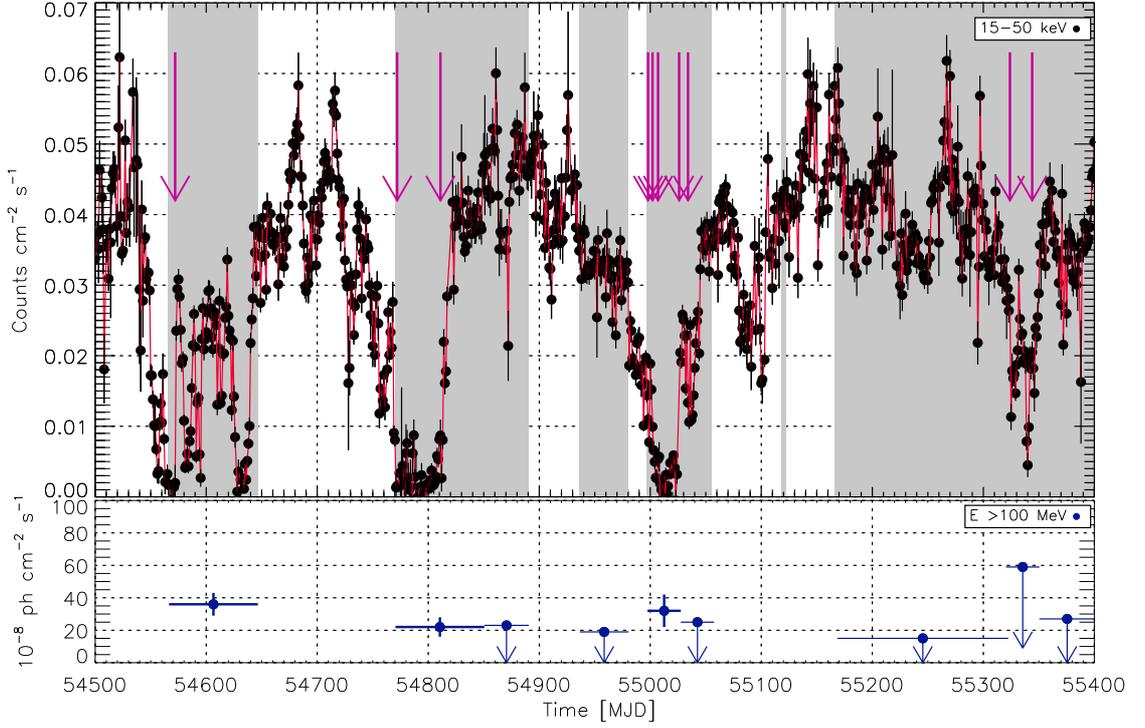}
\centering \caption { {\itshape (Top panel) \swift/BAT lightcurve (in counts per second in the energy range
15-50 keV) and AGILE-GRID \gray flares  for E$>$ 100 MeV as a function of time. The red arrows mark the dates
of major \gray flares of \s: MJD 54572, 54772, 54811, 54998, 55001, 55003, 55007, 55025, 55034, 55324, 55343.
Gray areas show the interval of good AGILE \gray exposure of \s. (Bottom panel) Average $\gamma$-ray flux
from  \s region in hard and soft X-ray state for E$>$100 MeV.  } }
\label{fig-1}
\end{figure*}

We detected several significant short timescale  $\gamma$-ray
flares from the \cyg  region during the 2009 Jun--Jul period
(pointing mode) and during the 2009 Dec.--2010 Jun period
(spinning mode). Fig.~\ref{fig-1} shows the  light-curve of \cyg
 obtained in the 15-50 keV range by the \swift-BAT instrument
from 2008 Jan. 1  to 2010 Jun 1. Superimposed on the same
plot are several arrows indicating the periods of major
$\gamma$-ray flaring detected by AGILE and associated with
\cygp, including those already reported in \citep{Tavani_2009b} (see Table \ref{table_A1}). 

Table \ref{table_3} and \ref{table_4} reports the details of the flares analyzed in this paper with the significance of the detections both with MSLA method and FDR method (when applicable). The MSLA and the FDRM use different assumptions for the
background (diffuse) model, therefore it could happen that the
relative significance of the detections is different. However, the
result we wanted to point out in the table is that both independent
methods would have selected the episodes as statistically significant.

Periods of long $\gamma$-ray AGILE exposure (shown in grey
in Fig.~\ref{fig-1}) cover both high, intermediate, and low states
of hard X-ray emission from \cygp. 

\subsection{Average $\gamma$-ray flux detected from \cyg}
\label{sec:average}
By integrating all AGILE data   from  \abe{2007 Jul 13} (MJD 54294.50) to \abe{2010 Jun 15} (MJD 55362.50)
(about 1.5  year of effective livetime on the Cygnus region) for E$>$ 100 MeV.
we find a $\gamma$-ray source at the $\sqrt{T_s} = 5.3$ level consistent with  \s position,
at (l,b) = (79.92, 0.58) $\pm$ 0.3 (stat.) $\pm$ 0.1 (syst.)
 and with an average flux  $F_1=(15 \pm 2) \cdot 10^{-8}$ \phcmsec above 100 MeV that contains
 all the $\gamma$-ray states. No significant detection is made for E$>$ 400 MeV.

If we remove all flares detected by AGILE with $\sqrt{T_s}
> 3$ (reported in \citep{Tavani_2009b} and in Table \ref{table_3}) we find
a $\gamma$-ray source \abh {at the $\sqrt{T_s} = 4.5$ level at the
position (l,b) = (80.03, 0.65) $\pm$ 0.5 (stat.) $\pm$ 0.1
(syst.), consistent with the \cyg position, with an average flux 
$F_2=(11.3 \pm 2.9) \cdot 10^{-8}$ \phcmsec above 100 MeV}. Note
that this extended time interval includes several periods with low
hard X-ray/high soft X-ray emission that we call here ``active
$\gamma$-ray states''. It is therefore possible that the value of
$F_2$ is influenced by low-level $\gamma$-ray emission from \cyg
during the active states. 

Table \ref{table_A1} reports all the  ``active'' and
``non-active'' $\gamma$-ray states associated with the hard and soft
X-rays state. These periods are also shown in the bottom panel of
Figure \ref{fig-1}. For example, integrating during the period 2009 Jun
15  until 2009 July 15 (MJD = 54997.50 - 55027.50)  we find
an average flux  $F_3=(32 \pm 10) \cdot 10^{-8}$ \phcmsec above
100 MeV. 

{\mt In order to study ``non-active $\gamma$-ray'' \cyg states
(i.e., those corresponding in general to the common high hard
X-ray states) we also focused on specific time intervals
characterized by average Swift/BAT levels of emission above 0.028
\ctscmsec. For the period starting on 2009 Jan. 19 until 2009 Feb. 28
 (MJD = 54850.50-54890.50)  we find a 95\% confidence limit
upper limit $F_4 < 23 \cdot 10^{-8}$ \phcmsec above 100 MeV.
Integrating from 2009 Dec. 1 until 2010 May 4 (MJD =
55166.50-55320.50) we find  a 95\% confidence limit upper
limit $F_5 < 15 \cdot 10^{-8}$ \phcmsec above 100 MeV. }

\begin {table}[!htb]
\caption {\em{Average $\gamma$-ray flux (E$>$ 100 MeV) from  \s region in hard and soft X-ray state.}}
\label{table_A1}
\renewcommand{\arraystretch}{1.2} % enlarge line spacing
\begin{tabular}{@{}llll}
\hline
 \textbf{ Time (MJD) } & $\sqrt{T_s}$ & Flux  & X-ray state\\
\hline
54566.50-54646.50 & 5.9 & 36 $\pm$ 7 & soft \\

54770.50-54850.50 & 3.6 & 22 $\pm$ 6 & soft \\

54850.50-54890.50 & $<$2  & $<$ 23 & hard\\

54936.50-54980.50 & $<$2  & $<$ 19 & hard \\

 54997.50-55027.50  & 3.5  & 32 $\pm$ 10  & soft \\

 55027.50-55055.50  &  $<$2  &  $<$ 20  & hard\\

 55166.50-55320.50 &  $<$2 & $<$ 15 & hard \\

 55320.50-55350.50 & $<$2 & $<$ 59 & soft intermediate \\

55350.50-55400.50 & $<$2 & $<$ 27 & hard \\
\hline
\end{tabular}
\vskip .1in
 The Table provides:  (1) the time interval in MJD; (2) the statistical significance
 \abc{$\sqrt{T_s}$} of the {source detection according to the maximum } likelihood ratio test;
 (3) the period-averaged flux $F$ (\enmev) in $10^{-8}$ \phcmsec (if $\sqrt{T_s}<3$ a $2-\sigma$
 upper limit is reported); (4) the X-ray state 
\end{table}

\subsection{Instrument stability}

\begin {table*}[!h]
\caption {\em{ $\gamma$-ray pulsars used to evaluate the instrument stability for E$\>$100 MeV.
}} \label{table_pulsar}
\renewcommand{\arraystretch}{1.2} % enlarge line spacing
\begin{tabular}{@{}llllllll}
\hline  
AGL source name  &  (l,b) positioning &  \gray flux  & $\sqrt{T_s}$  & Counterpart name  \\
                                     &  (degrees)   & ($10^{-8}$ \phcmsec) &      \\ 
  AGL J0247+6027   & (136.71,   0.70) $\pm$ 0.37  $\pm$ 0.1  & 11 $\pm$ 3 & 4.4 &   PSR J0248+6021    \\ 
  AGL J0303+7438   & (131.46, 13.98) $\pm$ 0.43  $\pm$ 0.1  & 7 $\pm$ 2 & 4.0 &   PSR J0308+7442     \\ 
  AGL J1136-6053    & (293.90,   0.67) $\pm$ 0.35  $\pm$ 0.1  & 12 $\pm$ 3 & 4.5 &   LAT PSR J1135-6055    \\ 
  AGL J1435-5932    & (315.68,   0.75) $\pm$ 0.70  $\pm$ 0.1  & 13 $\pm$ 4 & 4.0 &   LAT PSR J1429-5911    \\ 
  AGL J1953+3254   & (68.76,   2.86) $\pm$ 0.18 $\pm$ 0.1 & 17 $\pm$ 2 & 8.3 & PSR J1952+3252 \\
  AGL J2224+6113   & (106.20,   3.20) $\pm$ 0.43 $\pm$ 0.1 & 11 $\pm$ 2 & 5.7 & PSR J2229+6114 \\ 
  \hline
\end{tabular}
\end{table*}

\begin{figure}[!htb]
\centering
\includegraphics[width=9 cm]{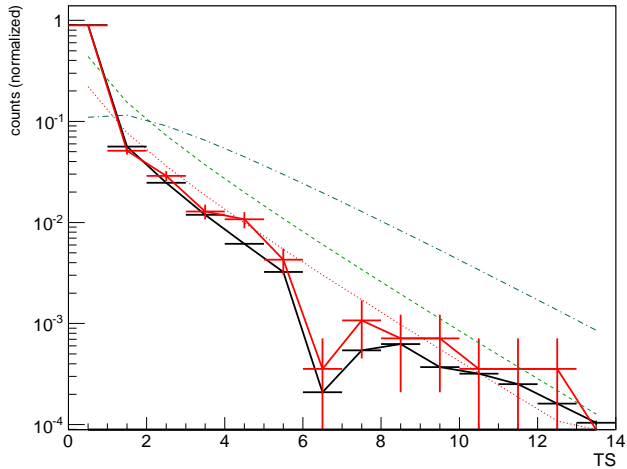}
\centering \caption { {\itshape
The red histogram is the probability density function (PDF) of $T_s$ values for real AGILE 
observations of the sources listed in Table \ref{table_pulsar}. 
The black histogram  is the PDF of the $T_s$ values of a simulated Cygnus region with all the sources reported 
in Table \ref{table_2}   except \cyg that has zero flux. The notch in the distribution near $T_s$=6 is caused by the switch between the fixed and the free position of the source in our analysis method. The red dotted line is the $\frac{1}{2}\chi^2_1$ theoretical
distribution, the green dashed line is the $\chi^2_1$  theoretical
distribution, the cyan dotted-dashed line is the $\frac{1}{2}
\chi^2_3$ distribution.} } \label{fig_TS}
\end{figure}

We have analyzed some AGILE sources having flux near the average $\gamma$-ray flux detected from \cyg (see Section \ref{sec:average}) that are not expected to be variable. In particular, we have considered the $\gamma$-ray pulsars reported in Table \ref{table_pulsar}.
The analysis has been carried out using the same procedure  performed on \cygp, by dividing the analyzed period (from 2007 Jul  to 2011 May) in two sets: pointing mode (with a 1 day bin size light-curve) and  spinning mode (from  2009 Nov., with a 2 days bin size light-curve). We have done about 1530 trials for the pointing mode and 1536 trials for the spinning mode. The resulting probability density function (PDF)  of $T_s$ values is shown in Fig. \ref{fig_TS} where it is compared with a simulation of \cyg field: in this field all the sources reported in Table \ref{table_2} are simulated except \cyg that has zero flux. It is noted that the $T_s$ distribution of real AGILE observations  of the source listed in Table \ref{table_pulsar} (red line) is fully compatible with the simulations (black line)   and this fact  excludes the presence of spurious flares introduced by instrument instability during the AGILE lifetime. As stated in \citep{Bulgarelli_2011} the vertical translation of the $T_s$ distribution depends on   the complexity of the analyzed region.  

With these distributions the expected number of wrong detections with $\sqrt(T_s) > 3.1$ (the lowest value of Table \ref{table_3}) in 150 maps (the number of bins of the light curves) is about 0.2; in this paper we have 8 detections in 150 maps.  

\subsection{Observations in pointing mode}

The AGILE pointing observations of the Cygnus field analyzed in
this paper are listed in Table \ref{table_1}, and cover about two
months of uninterrupted observation.
 Having obtained the general
results on the Cygnus region described in the previous section, we
carried out a systematic search for short timescale variability of
\cyg during the interval of substantial AGILE-GRID exposure in
pointing mode in 2009.

\begin {table*}[!h]
\caption {\em{ 1-day   \gray flares consistent with the \cyg
position reported in \cite{Tavani_2009b} for E$>$100 MeV.
}} \label{table_A1}
\renewcommand{\arraystretch}{1.2} % enlarge line spacing
\begin{tabular}{@{}llllllll}
\hline \gray flaring date  & (l,b) positioning &  \gray flux    & FDR-$\alpha$ & $\sqrt{T_s}$ \\
  (MJD)                                    &  (degrees)   & ($10^{-8}$ \phcmsec) &      \\ 
  54572.00-54573.00  & (79.10, 0.60) $\pm$ 0.60  $\pm$ 0.1  & 260 $\pm$ 80 & 0.001 & 4.2     \\ 
  54772.00-54773.00  & (79.30, 0.70) $\pm$ 0.70  $\pm$ 0.1  & 258 $\pm$ 82 & 0.01 & 4.0   \\ 
  54811.00-54821.00  & (79.60, 0.30) $\pm$ 0.60  $\pm$ 0.1  & 210 $\pm$ 73 & 0.01 & 3.8   \\ 
  \hline
\end{tabular}
\end{table*}

\begin {table*}[!h]
\caption {\em{Main 1-day and 2-days \gray flares consistent with the \cyg
position detected in 2009 Jun--Jul  and  from 2009 Dec. to 2010 Jun for E$>$100 MeV. The last row reports the sum of all the flares with the repeated flare occurrence post trial significance.
}} \label{table_3}
\renewcommand{\arraystretch}{1.2} % enlarge line spacing
\begin{tabular}{@{}llllllll}
\hline \gray flaring date  & (l,b) positioning &  \gray flux  & FDR-$\alpha$ & $\sqrt{T_s}$  & pre-trial $\sigma$ & post-trial$\odot$ $\sigma$ & \swift-BAT flux \\
  (MJD)                                    &  (degrees)   & ($10^{-8}$ \phcmsec) & & &for a single   & for a single   & ($cts cm^{-2} s^{-1}$) \\
      &     &   & & &  detection   &   detection  &  \\
 \hline
 54997.50-54998.50  & (80.50, 0.53) $\pm$ 0.90  $\pm$ 0.1  & 180 $\pm$ 64 & 0.005 & 3.9 & 3.75  &  2.2 & 0.010 \\
 55001.00-55002.00   & (79.44, 0.93) $\pm$ 0.65  $\pm$ 0.1  & 168 $\pm$ 67 & $10^{-5}$  & 3.5 & 3.35  & 1.57  & 0.015 \\

 55003.00-55004.00  & (80.35,   1.15) $\pm$ 0.64  $\pm$ 0.1  & 157  $\pm$ 56 & $10^{-5}$ & 3.8  & 3.71  &  2.11 & 0.007 \\

 55007.00-55008.00  & (79.30, 0.81) $\pm$ 0.57  $\pm$ 0.1  & 176 $\pm$ 64 & $10^{-5}$ & 3.7  & 3.51  & 1.83  & 0.005 \\
 55025.00-55026.00  & (79.75, 1.15) $\pm$ 0.58  $\pm$ 0.1  & 167 $\pm$ 70 & 0.006 & 3.3  & 3.20  &  1.29 & 0.008 \\
 55034.00-55035.00  & (80.12, 0.96) $\pm$ 1.00 $\pm$ 0.1  & 168 $\pm$ 67 & 0.005 & 3.4  &  3.27 & 1.43  & 0.013 \\
 
 55324.00-55326.00* & (79.12, 0.91) $\pm$ 0.83  $\pm$ 0.1  & 170 $\pm$ 70 & - & 3.3  &  3.20 &  1.29 & 0.028 to 0.011 \\
 55343.00-55345.00$\dagger$ & (79.89, 0.71) $\pm$ 0.86 $\pm$ 0.1  & 290 $\pm$ 103 & - & 3.7  &  3.51 & 1.83  & 0.020 to 0.018 \\  \hline
   &    &   &    &  & & repeated flare   \\
     &    &   &    &  & & occurrence    \\
    &    &   &    &  & & post-trial  $\sigma$ \\
\hline
 Sum of above detections  & (79.69, 0.72) $\pm$ 0.30 $\pm$ 0.1  & 160 $\pm$ 40 &  & 6.2  &   & 7.2 \\
\hline
\end{tabular}
\vskip .1in
* ATel 2609.\\ $\dagger$  ATel 2645 \\ $\odot$ We calculated the post-trial significance for 150 trials, which is equivalent of about 1 year of total exposure time.
\end{table*}

\begin {table*}[!h]
\caption {\em{ Main 1-day and 2-days \gray flares consistent with the \cyg
position detected in 2009 Jun--Jul  and  from 2009 Dec. to 2010 Jun. for  E $>$ 400 MeV.  The last row reports the repeated flare occurrence post-trial significance.
}} \label{table_4}
\renewcommand{\arraystretch}{1.2} % enlarge line spacing
\begin{tabular}{@{}llllllll}
\hline
\gray flaring date  & (l,b) positioning &  \gray flux  & $\sqrt{T_s}$  & pre-trial $\sigma$& post-trial $\sigma$  & \swift-BAT flux \\ 
                   (MJD)                   &  (degrees)   & ($10^{-8}$ \phcmsec) & & for a single & for a single   & ($cts cm^{-2} s^{-1}$) \\
                    &     &   &  &  detection   &   detection  &  \\
\hline
 54997.50-54998.50 & (80.42, 0.41) $\pm$ 0.60 $\pm$ 0.1 & 67 $\pm$ 29 & 3.9   &  3.75 & 2.2  & 0.010 \\
 55000.50-55001.50 & (78.88, 0.56) $\pm$ 0.92 $\pm$ 0.1 & 45 $\pm$ 24 & 3.0  & 2.99 & 0.89 & 0.015 \\
 55018.50-55019.50 & (79.75, 0.55) $\pm$ 0.60 $\pm$ 0.1 & 65 $\pm$ 33 &  3.7  & 3.51 & 1.83 & 0.004 \\
 55343.50-55345.50 & (80.07, 0.56) $\pm$ 0.70 $\pm$ 0.1 & 94 $\pm$ 43 &  3.5  &  3.35 & 1.57  & 0.020 to 0.018 \\
\hline
   &    &   &    &  & repeated flare  & \\
     &    &   &    &  & occurrence  & \\
    &    &   &    &  &  post-trial $\sigma$ & \\
\hline
   Sum of above detections  & (79.8, 0.5)  $\pm$ 0.31 $\pm$ 0.1  &  58 $\pm$ 18 & 5.1   &  & 3.8 & \\
\hline
\end{tabular}
\vskip .1in
\end{table*}

\begin{figure*}[!htb]
\centering
\includegraphics[width=18cm]{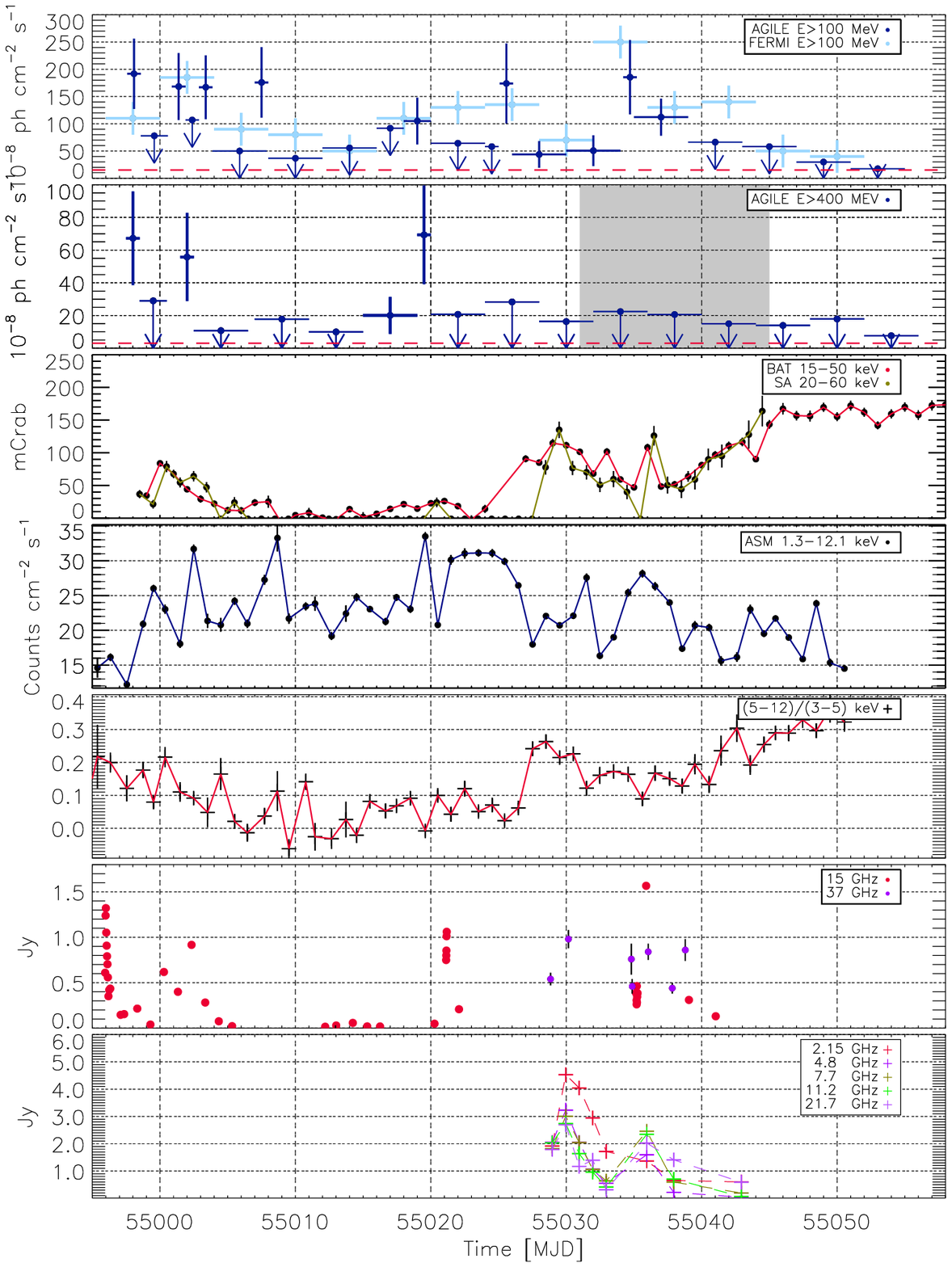}
\centering \caption { {\itshape AGILE/GRID, AGILE/SA, Fermi/LAT, Swift/BAT,  XTE/ASM, AMI-LA, RATAN-600 and
Mets{\"a}hovi Radio Observatory data of \s during the uninterrupted 2-months period 2009 Jun.--Jul.
(Panel 1 (top)) the AGILE-GRID \gray light curve with a variable window time to put in evidence the \cyg
flares of 1-day timescale, the other data and upper-limits are determined with 1- 2- or 4-day time intervals;
the Fermi light curve of 4-days timescale. E$>$100 MeV.
(Panel 2) the AGILE-GRID \gray light curve for E$>$400 MeV and the coverage of MAGIC observation (gray area).
(Panel 3) the hard X-ray light curve as monitored by BAT   (15-50 keV) and by Super-AGILE (20-60 keV) with a
daily timescale bin.
(Panel 4) the soft X-ray light curve  as monitored by XTE-ASM (1.3-12.1 keV) for a 1-day integration time bin;
(Panel 5) the ASM hardness ratio $(5-12 keV)/(3-5 keV)$ data.
(Panel 6) AMI-LA  radio flux monitoring of \cyg  at 15 GHz and the  Mets{\"a}hovi Radio Observatory data at
 37 GHz.
(Panel 7) RATAN-600 radio telescope data at different frequencies.
The AGILE/GRID $\gamma$-ray upper limits are at the $2-\sigma$ level, the flux error bars are $1-\sigma$
values. } }
\label{fig-B}
\end{figure*}

Fig.~\ref{fig-B}, panels 1-3 provide a close-up of Fig.~\ref{fig-1} showing the details of the $\gamma$-ray
emission above 100 MeV (panel 1) and 400 MeV (panel 2) detected by AGILE from \cyg together with simultaneous
hard X-ray information from both \swift-BAT and Super-AGILE (panel 3). The interval covering the
period 2009 Jun.--Aug. (MJD: 54997.50 - 55055.50) corresponds to a
deep and prolonged minimum of the hard X-ray emission from
\cygp. The Fermi light curve is also superimposed in panel 1 \cite{Abdo_2009b}.
Several 1-day episodes of $\gamma$-ray flaring emission from \cyg
 with flux $F > 100 \times 10^{-8} \, \rm$ $\phcmsec$ are detected. \abe{ Table \ref{table_3} provides detailed information on  the main flare episodes
of Fig.~\ref{fig-B}  and obtained for 1-day time integrations  for $E>100$ MeV. Table \ref{table_4} lists  flares for $E>400$ MeV. 
The coverage of one of the MAGIC observations \citep{magic}  of \cyg in the 
soft X-ray state that provides an upper limit above 250 GeV is also shown in panel 2.

}

Fig.~\ref{fig-B}, panel 5-6  shows the radio monitoring data obtained
by our group  at
15 GHz at the AMI-LA radio telescope (already published in \cite{Abdo_2009b}), the RATAN-600
data at the frequencies 2.15, 4.8, 7.7,
11.2, and 21.7 GHz, and the Mets{\"a}hovi Radio Observatory data at 37 GHz. 
Finally, Fig.~\ref{fig-B}, panels 4-5 shows  the
1.3-12 keV data obtained from the XTE/ASM monitoring of \cygp.

\begin{figure*}[!htb]
\centering
\includegraphics[width=12cm]{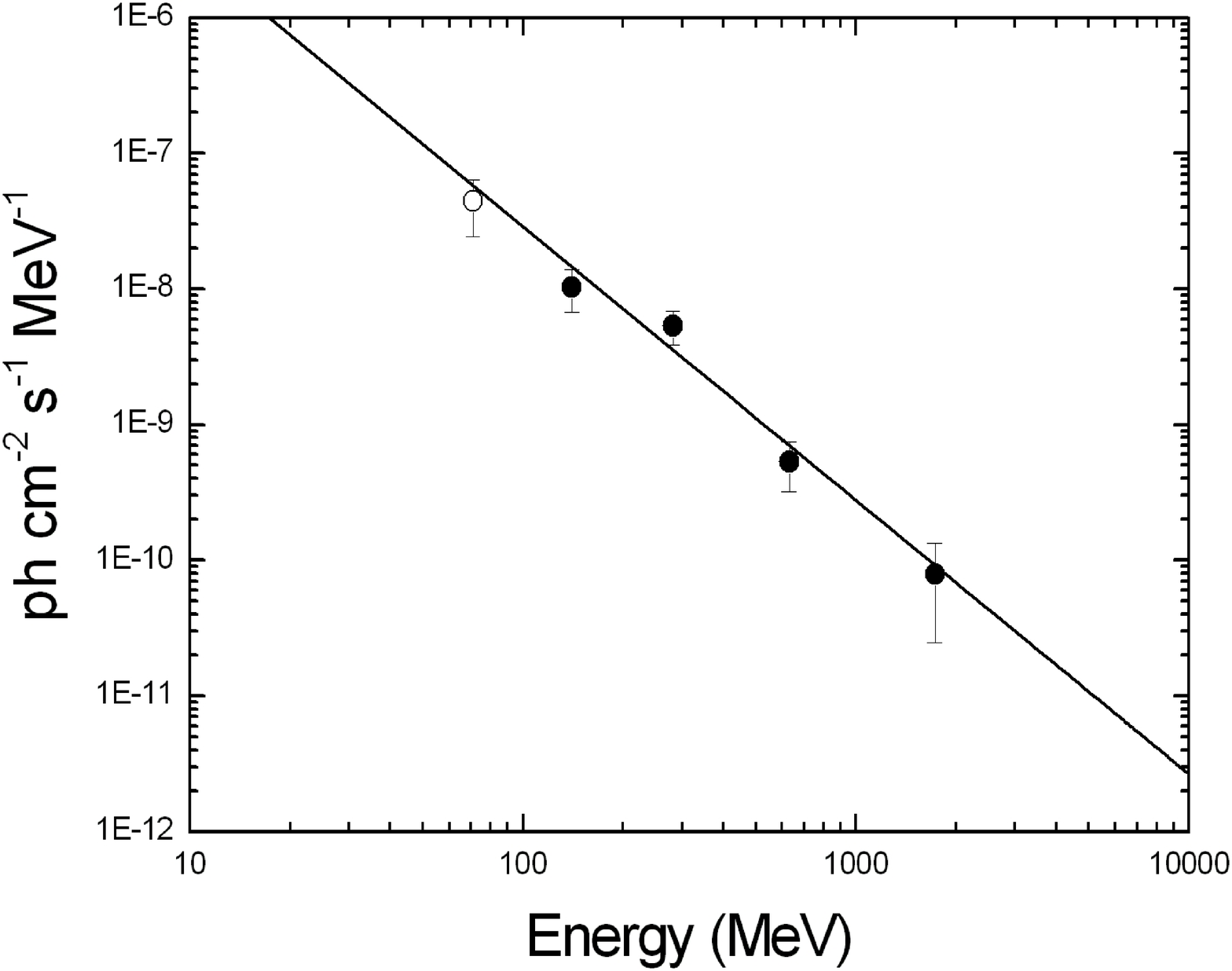}
\centering \caption { {\itshape  The AGILE-GRID $\gamma$-ray spectrum of \cyg
obtained by integrating all flaring episodes in 2009 Jun.--Jul. A single power-law
model fitting gives a photon spectral index is 2.02 $\pm$ 0.28. \aba{The open circle indicates the energy
channel not used for fitting}}.}
\label{fig-5}
\end{figure*}

Fig.~\ref{fig-5} shows the $\gamma$-ray spectrum obtained by integrating all of the
$\gamma$-ray flaring episodes in \abe{2009 Jun.--Jul}. A single power-law fit gives a photon
spectral index $\alpha = 2.0 \pm 0.2$. However, a more complex spectrum with
substantial curvature in the energy range 100 MeV - a few GeV cannot be
excluded.

\subsection{Observations in ``spinning'' mode}

Since early 2009 Nov. the AGILE satellite \aba{has been} operating in a
``spinning'' mode implying a smooth and continuous change of the
satellite attitude. In this mode, the $\gamma$-ray boresight axis
sweeps 360 degrees in about seven minutes. Solar panels are kept
perpendicular to the Sun direction by an automatic mechanism,
so that the pattern swept on the sky slowly moves with
time following the solar panel configuration. The resulting
$\gamma$-ray daily exposure covers about 70\%
of the whole sky every day (leaving uncovered only the regions
near the Sun or anti-Sun directions) and  provides significant
continuous monitoring of exposed sources for many months.
The Cygnus region was in a  favorable position in
the sky, i.e., always in the field of view (within about 50
degrees off-axis) of the AGILE spinning instrument. This was useful for assessing in a consistent way
the pattern of
$\gamma$-ray variable emission from \cyg  and correlating with other
wavelength emission. The AGILE pointing mode had a relatively
larger daily exposure compared with that obtained in the spinning
mode. However, the AGILE pointings at the Cygnus region required a
pre-defined planning strategy, and the overall monitoring livetime
was about 50\% from 2007 Jul to
2009 Oct.

Since AGILE started operating in spinning mode
(2009 Nov.) to 2010 Jun. \cyg has been mostly in its  
hard X-ray state that does not favour strong $\gamma$-ray emission \citep{Tavani_2009b, Abdo_2009b} .

We continuously searched for transient $\gamma$-ray emission from the
\cyg region during the AGILE spinning mode phase with an automatic alert monitoring system
\citep{Bulgarelli_2009}. In 2010 May this monitoring
 system  detected a  signal above $\sqrt(T_s)=3.5$
from \cygp, that was subsequently verified by a manual analysis.
We detected two $\gamma$-ray emission episodes during MJD 55324-55326
\citep{atel2609} not confirmed by Fermi \citep{corbel2010a}, and  on MJD 55343-55345
\citep{atel2645} confirmed by Fermi \citep{atel2646}.
The details of a refined analysis are reported in Table
\ref{table_3}  for $E>100$ MeV and in Table \ref{table_4} for E$>$400 MeV. 

\begin{figure*}[!htb]
\centering
 \includegraphics[width=12cm]{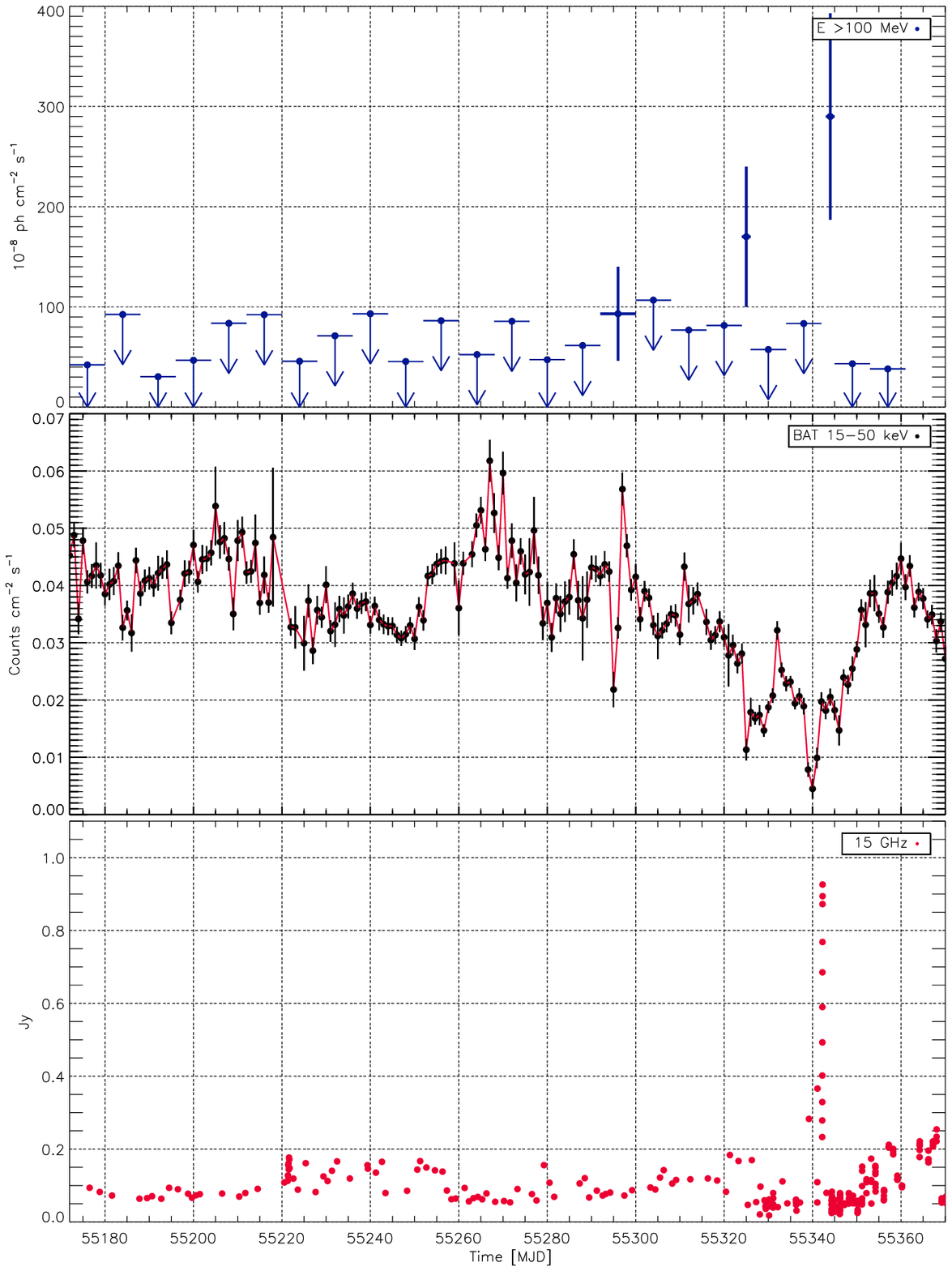}
% \resizebox{\hsize}{!}{\includegraphics{position.jpg}}
\centering \caption { {\itshape Simultaneous AGILE/GRID, \swift/BAT and radio monitoring data of \s during
the uninterrupted 18-months in the period 2009 Dec.--2011 May.
(Top panel:) the AGILE-GRID \gray lightcurve for E$>$100 MeV.
(Second panel:) \swift/BAT lightcurve (in counts per second in the energy range 15-50 keV).
(Third panel:) AMI-LA  radio flux monitoring of \cyg  at 15 GHz.} }
\label{fig_1_2010}
\end{figure*}

Fig.~\ref{fig_1_2010}  shows the detailed 1-day hard X-ray lightcurve of \cyg together
with the AGILE-GRID emission episodes and AMI-LA radio monitoring. 
A 15 GHz radio flare is evident in the data and it anticipates  the $\gamma$-ray flare at MJD 55343 by $\sim$1-day.

\subsection{Combining all Cygnus X-3 flares detected by AGILE}
\label{sec:comb}

\begin{figure*}[!htb]
\centering
\includegraphics[width=12cm]{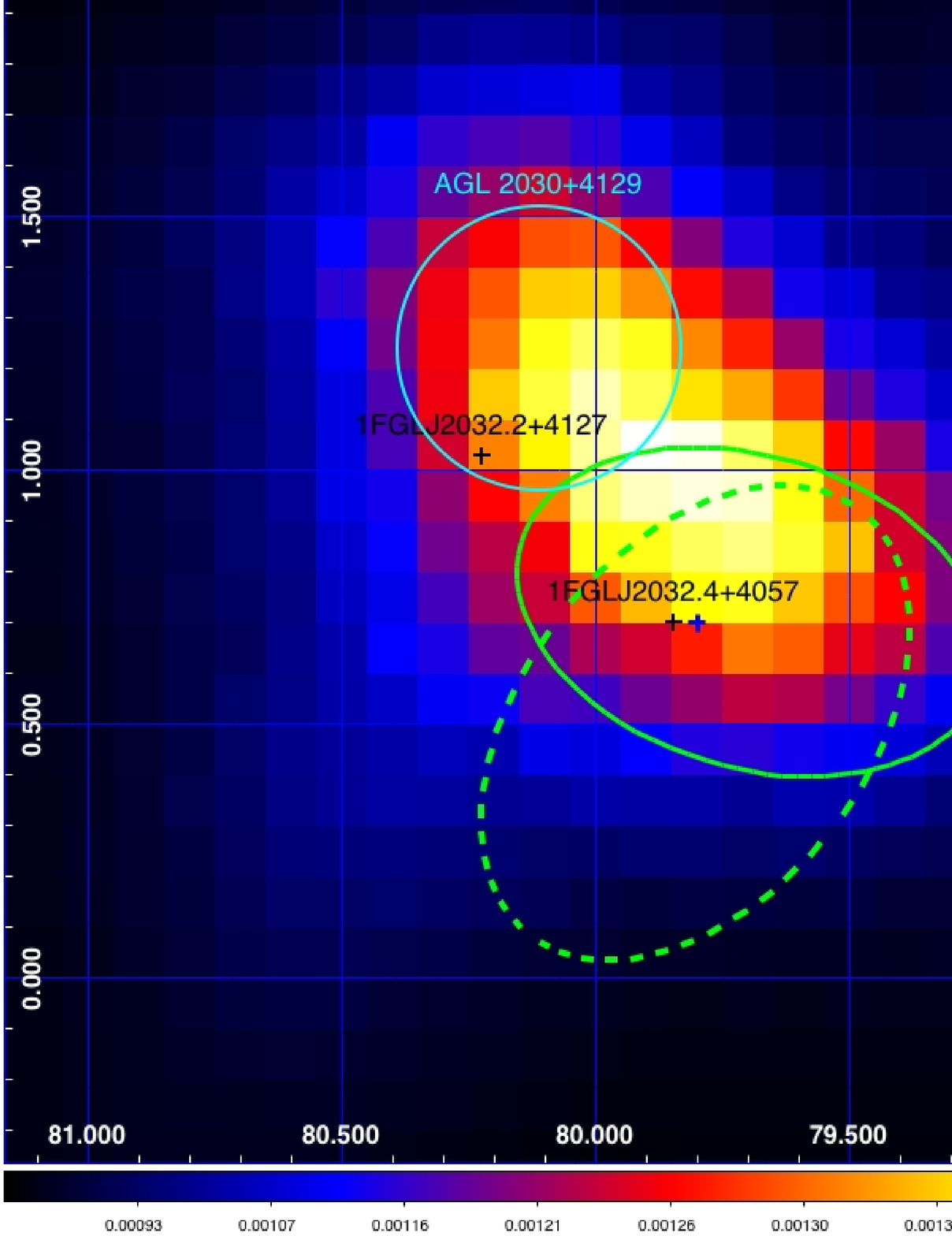}
\centering \caption { {\itshape  The $\gamma$-ray intensity map of the sum of all the \cyg flares reported in this paper, in Galactic coordinates for E$>$100 MeV. Pixel size=0.1$\degmark$ with  5-pixel Gaussian smoothing. Cyan contours: AGILE-GRID 95$\%$ confidence level of LAT PSR J2032+4127; Green continuous contour: the \cyg flares 95$\%$ confidence level (statistical and systematic errors) for E$>$100 MeV. Green dashed contour: the \cyg flares 95$\%$ confidence level (statistical and systematic errors) for E$>$400 MeV. Black crosses: Fermi-LAT (1-year catalog) sources (crosses shown for contours too small to be visible). The green contour has been calculated with the multi-source likelihood analysis method described in the text, using persistent sources reported in Table \ref{table_2}. Blue cross is the position of \cygp.} }
\label{fig-ffs}
\end{figure*}

Fig. \ref{fig-ffs} reports the sum of all the flares
detected by AGILE-GRID from \cyg reported in this paper for E$>$100 MeV. The
significance of the sum of all the flares is $\sqrt{T_s} = 6.2$
with a flux of 160 $\pm$ 40 \phcmsec and a 95$\%$ contour level centered in
$(l,b)=(79.69, 0.72)$ in Galactic coordinates, with a semi-major
axis of 0.47$\degmark$ and a semi-minor axis of 0.28$\degmark$
(including statistical and systematic errors). 

The significance of the sum of all the flares for E$>$400 MeV is $\sqrt{T_s} = 5.1$
with a flux of 58 $\pm$ 18 \phcmsec and a 95$\%$ contour level centered in
$(l,b)=(79.8, 0.5)$ in Galactic coordinates, with a semi-major
axis of 0.53$\degmark$ and a semi-minor axis of 0.34$\degmark$
(including statistical and systematic errors).

The 1FGL~J2032.2+4127 and  AGL~2030+4129 are both outside the AGILE
error box. This fact excludes that the detected flares
originate from these nearby sources. 

{\mt In order to assess the statistical significance of our
detections, we consider the post-trial probability of flare
occurrence. We have to distinguish two cases:

\begin{enumerate}
\item  the case of a single flare episode originating from a specific source within a given error box 
(that we define as ``single independent occurrence'');
\item
the case of repeated flaring episodes originating from a specific
source with a given error box  (that we call here ``repeated flare
occurrence''.
\end{enumerate}
For each individual detection by AGILE reported in Tables
\ref{table_3} and \ref{table_4} 
we calculated the
post-trial significance of the \textit{single independent
occurrences},  which does not take into account the
history of repeated occurrences.}

\abh{ We calculate the post-trial significance for
\textit{repeated flare occurrences} at the \cyg error-box
position as follows. Each independent time period is a single trial. The chance probability
of
having $k$ or more detections at a specific site with a $T_s$
statistic satisfying $T_s \ge h$ in $N$ trials is given by $ P(N, k)
= 1 - \sum_{j=0}^{k-1}
\left(\begin{array}{c}N\\j\end{array}\right) p^j (1-p)^{N-j}$
where $p$ is the p-value corresponding to the $h$ value. For $T_s
\ge 10.9$, we have a p-value of $6.8 \times 10^{-4}$. The estimated
probability of 8  detections (E$>$100 MeV) consistent with the null hypothesis
in 150 maps is  $P = 2.3792 \times 10^{-13}$ that
corresponds to $\sim7.2$ Gaussian standard deviations.}

Table \ref{table_4} reports the repeated flare occurrence significance for E$>$400 MeV.

In order to test the pre-trial type I error (rejecting the null hypothesis when in fact it is true) we also searched for transient emission at the position (l,b)=(78.38, 0) near \cyg with similar characteristics of the diffuse background. We found a $T_s$ distribution compatible with the expected statistics (the black histogram of Fig. \ref{fig_TS}).

%%%%%%%%%%%%%%%%%%%%%%%%%%%%%%%%%%%%%%%%%%%%%%%%%%%%%%%%%%%%%%%%%%%%%
\subsection{Statistical correlations with other wavelenghts}
%%%%%%%%%%%%%%%%%%%%%%%%%%%%%%%%%%%%%%%%%%%%%%%%%%%%%%%%%%%%%%%%%%%%%
\label{sec:correl}

\subsubsection{Anticorrelation with hard X-ray emission}

We notice that the $\gamma$-ray episodes detected by AGILE are
all in correspondence with peculiar states of the hard X-ray
emission. Both the $\gamma$-ray emission flaring episodes
 correspond to   $\sim$1-day
minima of the BAT hard X-ray light-curve. The typical trend of
anticorrelated soft and hard X-ray emission from \cyg is evident and also shown by the ASM
hardness ratio trend.

This paper reports eight flares originating from \cyg for E$>$100
MeV. All are in correspondence with a low hard X-ray flux of \cygp; 
seven of them are under 0.02 \ctscmsec in the Swift/BAT, and one
is under 0.028 \ctscmsec.  To quantify the relation between
$\gamma$-ray emission and  hard X-ray emission  it is
reasonable to set  a threshold of 0.028 \ctscmsec in the Swift/BAT
data: the fraction of the time that this
source is under this level during the AGILE observation time is
$f=0.32$.  For $k$ $\gamma$-ray flares
originating from \cygp, $f^k$ is the probability that
all the flares are under the hard X-ray threshold, and $P(N,k)$ is the
probability of having $k$ or more detections in $N$ maps (see
Section \ref{sec:comb}). We find that the  confidence level that
all the $\gamma$-ray flares occur at a low hard X-ray flux is $>
7.7$ normal standard deviations.

We have also calculated the Discrete Cross-Correlation Function (DCF).  
The DCF correlates two sets of unevenly sampled data
\citep{Edelson_1988}. We have correlated both BAT/Swift (sampled each day) and AMI-LA
(variously sampled more than one time in a day or every few days) with AGILE data
(sampled every few days, with a variable time bin size depending by the $\gamma$-ray state
of the source).
The calculation of the DCF uses a DCF binning of 8 days to take into account the
biggest time bin size of AGILE light curve, with a scan time lag of 1 day.
The UL are converted into flux and related 1$\sigma$ error considering half of the upper limit.
The significance level of the peak was determined by simulating  AGILE light curves
with a bootstrap selection from the original points.

Choosing the right DCF bin size is a trade off between the high accuracy of the correlation
coefficient between sets of data of two time series and the resolution in the description of the
 cross-correlation curve. We have performed tests with different bin sizes of the AGILE light curve
 (splitting the same flux in smaller bins) and we have tried also different DCF bin sizes with the
 AGILE original light curve of Figure \ref{fig-B} and \ref{fig_1_2010}. In any case we have found
 that the peak of the DCF depends only weakly on the specific bin size. Due to the bin size of AGILE
 light curve, we use the DCF only for the determination of the cross-correlation and not for a measurement of a lag 
 between hard-X and $\gamma$-rays.
 
 Correlating hard X-rays and $\gamma$-rays we find a negative correlation with a 3$\sigma$ confidence
level between these frequencies compatible with 0 $\pm$ 8 days lag.

\subsubsection{Correlation with hard X-ray emission}

Despite some gaps in the radio data, in particular during the pointing mode observations, several radio flares are evident in the data. We find a positive correlation above the 3$\sigma$ confidence level between the radio and $\gamma$-rays with a lag of 0 $\pm$ 8 days. The same consideration reported for the DCF between X-rays and $\gamma$-rays is valid for the correlation between the radio and $\gamma$-rays, and for this reason the DCF is not used for a measurement of the lag.

\section{Discussion}

The AGILE $\gamma$-ray monitoring of  \cyg contributes in a
significant way to the determination of a repetitive pattern of
$\gamma$-ray emission and particle acceleration in this microquasar.
Several fast (1- or 2-day) $\gamma$-ray flaring episodes are
detected in soft X-ray states. Furthermore, persistent and
significant $\gamma$-ray emission is detected in AGILE data
usually during a low flux in hard X-rays.
 The  most intense $\gamma$-ray emission above 100 MeV
distinctively occurs in coincidence with minima of the hard X-ray
emission. As we discuss below, on several occasions a radio flare
was detected near or after the $\gamma$-ray peak emission.
This trend, that was already apparent  in \citep{Tavani_2009b} and
\citep{Abdo_2009b}, is confirmed by a more extended monitoring
over more than 1 year.

The $\gamma$-ray flaring episodes are demonstrated to be even more relevant
for understanding the source dynamics when compared to soft X-ray, hard X-ray, and
radio data. Fig. \ref{fig-B} and \ref{fig_1_2010}  show the $\gamma$-ray datapoints
with multifrequency information.

\subsection{General characteristics}

{\mt Observationally, $\gamma$-ray emission from \cyg  starts to
be detectable by AGILE} when the \swift-BAT hard X-ray flux 
 decreases below a count rate flux of $F = 0.02$ \ctscmsec, as shown by the data reported
in Table \ref{table_3} and \ref{table_4} 
 (see Fig.
\ref{fig-B} and \ref{fig_1_2010}). Furthermore, $\gamma$-ray
flaring appears to be occurring during or immediately before
sudden X-ray spectral transitions.
The $\gamma$-ray flare recorded on
MJD~55324  is associated with a spectral transition from hard
to soft X-ray emission, although unlike the other detections, it is not coincident with a radio flare.

The reported data  show in general an anti-correlation
between the $\gamma$ and hard X-ray fluxes, and a correlation
between the $\gamma$ and soft X-ray fluxes\footnote{\abe{We notice
that occasionally during rapid variations of the X-ray flux the
$\gamma$-ray activity increases and becomes detectable at a
$\sqrt{T_s} > 3$ even outside the soft X-ray state. See the case
reported for MJD 55292-55300 and 55362-55370.}.}.

 Several $\gamma$-ray flares (55001, 55019, 55034) are
associated with \textit{1-2 day delayed} radio flares at the
level of 1 Jy at 15 GHz as detected by AMI-LA. For other $\gamma$-ray
flares, the relation with the radio emission is not so obvious,
either for lack of radio monitoring (see the episode at
MJD~55025), or for the quasi-simultaneous $\gamma$-ray and radio
emission (as in  MJD~55343, see Fig.~4).
However, the majority of \cyg  $\gamma$-ray flares reported here
and in \citep{Tavani_2009b} appear to follow a trend characterized by a
substantial $\gamma$-ray enhancement \textit{preceding} radio flaring activity by a few days, which is within the
formal DCF error bar of 8 days.
The $\gamma$-ray flare on MJD 55019 is possibly associated
with a $\gamma$-ray harder spectrum compared to other flares. 

 An interesting case is provided by the event near  MJD 55025.
 A prominent radio flare reaching 4 Jy at 2.1 GHz (RATAN-600) and 1
Jy at 37 GHz (Mets{\"a}hovi) occurred on MJD 55030, followed by a
secondary \abc{radio flare with inverted spectrum} near MJD 55035
detected by RATAN-600  as well as by AMI-LA and Mets{\"a}hovi.
AGILE data show  that a 1-day $\gamma$-ray flare was detected on
MJD 55025, i.e., almost 5 days before the strong radio flare. Lack
of radio monitoring preceding this flare does not allow us to
determine the association of the MJD 55025 $\gamma$-ray flare with
other radio flaring activity. In any case, during the $\sim$4 days
between the $\gamma$-ray and radio flares the overall $\gamma$-ray
flux stayed at a low level. In particular, during the radio flare
itself that lasted about 3 days, the $\gamma$-ray flux was below
$F = 100 \rm \; \times 10^{-8}  \; ph. cm^{-2} \,  s^{-1}$ near
100 MeV and above.

A  sequence of $\gamma$-ray emission immediately followed.
On MJD 55034-55035 the $\gamma$-ray flux shows an increase above $F =
180 \rm \, \times 10^{-8}  \; ph. cm^{-2} \,  s^{-1}$ \abm{for E$>$100 MeV} followed by  4 days of high
level flux and subsequent  decay. One day later (MJD 55035 and
following days), the ``secondary'' radio flare with the inverted
spectrum was detected reaching 2 Jy at 11.2 GHz.
The excellent multi-frequency coverage by our group
  during this period (MJD 55030-55040) is enhanced by
 several TeV observations of \cyg by the MAGIC group  \cite{magic} followed after an AGILE alert.
 It is then
 interesting to note that the MAGIC 95\% confidence upper limit of
 $4 \times 10^ {-12} \rm \, ph. cm^{-2} \,  s^{-1}$ above 250 GeV obtained
 from a set of observations including those of interest here provide, for the
 first time for \cygp,  simultaneous broad-band spectral information from 100 MeV to TeV
 energies.

The AMI-LA radio data of MJD 55343 (see Fig. \ref{fig_1_2010}) shows the correlation between radio
and $\gamma$-ray flares with the radio flares that anticipate the $\gamma$-ray flare. The same behaviour
is reported in \citep{Williams_2011}.

\subsection{The $\gamma$-ray spectrum}
\label{sec:spectra}
Fig. \ref{fig-5} shows the $\gamma$-ray spectrum  obtained by summing all the major
aforementioned episodes during the period 2009 Jun.-Jul. The complex relation between
the $\gamma$-ray emission of \cyg near 100 MeV up to 1 GeV and the X-ray and radio spectrum
will be addressed in detail elsewhere.  We emphasize here that the AGILE results
set interesting constraints on the hardness of the $\gamma$-ray spectrum near 100 MeV.
Our measured power-law spectrum is consistent in being ``flat'', that is with a power-law
index near 2 in the AGILE-GRID energy range. However, we cannot exclude the
existence of spectral  curvature of the $\nu F_{\nu}$ spectrum in the energy range
below or near 100 MeV. We address this point as  well as the analysis and
implications  of the measured broad-band spectrum of \cyg in  \citep{Piano_2011}.

\subsection{AGILE and Fermi comparison}

We compare our results on \cyg with those of \fermi-LAT. There is
considerable overlap of emission episodes detected by the two missions during the
period 2009 Jun.-Jul. Fig. \ref{fig-B} (top panel) shows the AGILE $\gamma$-ray lightcurves
of \cyg with different time bins and the available \fermi-LAT $\gamma$-ray lightcurve \citep{Abdo_2009b}.
The two instruments have a quite different response and daily exposure at energies
near 100 MeV, and the daily effective exposure is influenced by solar panel and other
geometric constraints of the pointing strategy. Furthermore, for a source of relatively
rapidly varying (within one day or shorter timescale) $\gamma$-ray emission such as \cygp,
the two instruments can catch different states of emission. \abc{The} 1-day flaring AGILE-GRID
lightcurve generally agrees with the 4-day averaged \fermi-LAT lightcurve.
Panel 2 of Fig. \ref{fig-B} shows the MJD 55019 hard $\gamma$-ray
episode detected by AGILE (associated with the radio flare of MJD 55021 and ASM X-ray peak)
which is not evident in the 4-day average of the corresponding \fermi-LAT data shown in the top panel.

\begin{figure*}[!htb]
\centering
\includegraphics[width=12cm]{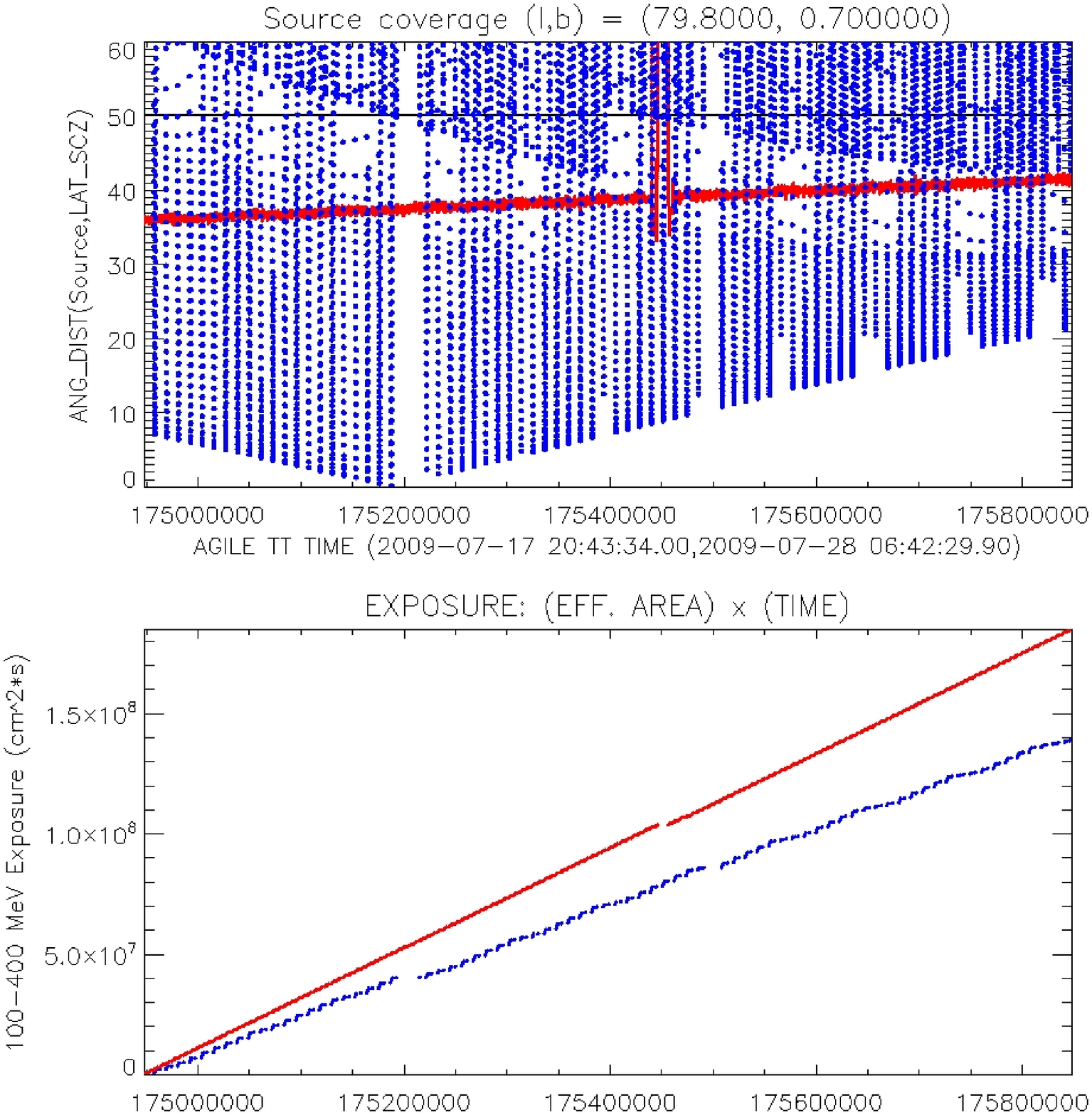}
% \resizebox{\hsize}{!}{\includegraphics{position.jpg}}
\centering \caption { {\itshape  \mt  Different pointing
strategies of \s for AGILE-GRID (in red) and \fermi-LAT (in blue).
\textit{Upper panel:} time evolution of the  instrument off-axis
angle with respect to the fixed \cyg position during the
period 2009 Jul 18-28 \abh{when the AGILE operates in pointing mode}. The curve in blue  show the off-axis
angle evolution for the all-sky scanning pointing strategy of
\fermi-LAT. The curve in red  show the off-axis angle for the
AGILE-GRID fixed pointing strategy in 2009 Jul. \textit{Bottom panel: }
time evolution of the cumulative exposure between 100 MeV and 400 MeV for \fermi-LAT
(blue curve, \abh{using public data and Pass 6 Version 3 Front photons Instrument Response Function}) and AGILE-GRID (red curve) assuming a starting time
on 2009 Jul 18. } } \label{fig-11}
\end{figure*}

\begin{figure*}[!htb]
\centering
\includegraphics[width=12cm]{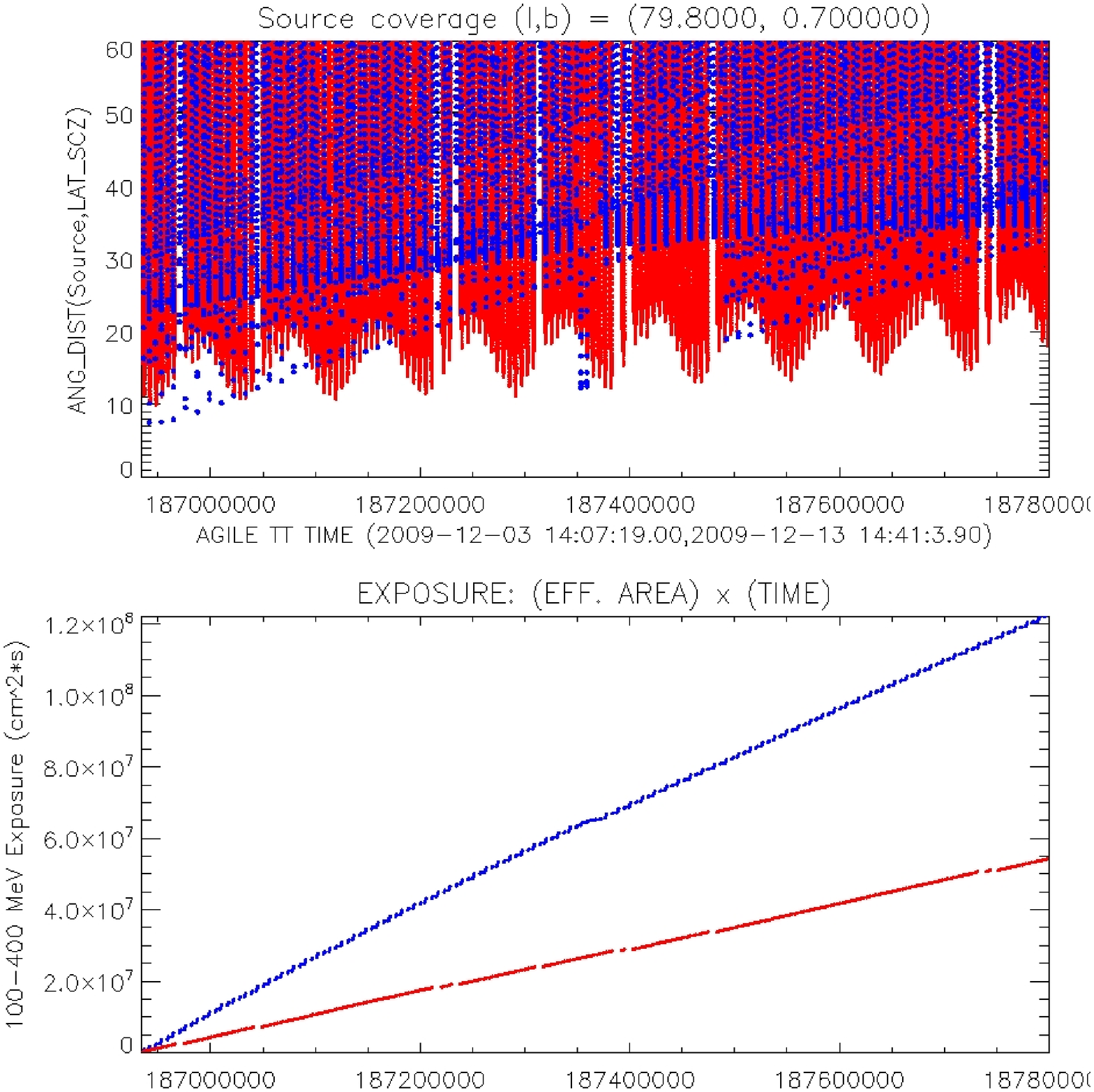}
% \resizebox{\hsize}{!}{\includegraphics{position.jpg}}
\centering \caption { {\itshape  Same pointing strategies of \s for AGILE-GRID (in red) and
Fermi-LAT (in blue). \textit{Upper panel:} time evolution of the instrument
off-axis angle with respect to the fixed \s position during the
period 2009 Dec. 3-13, when AGILE started to operate in spinning
mode.The curves show the off-axis angle evolution for the all-sky
scanning pointing strategy of AGILE GRID (in red) and \fermi-LAT (in
blue). \textit{Bottom panel:} time evolution of the cumulative exposure between
100
MeV and 400 MeV for Fermi-LAT (blue curve, using public data and Pass 6 Version 3 Front photons Instrument Response Function) and AGILE-GRID (red curve)
assuming a starting time on 2009 Dec. 3.} }  \label{fig-12}
\end{figure*}

Fig. \ref{fig-11} (top panel) shows the time evolution of the off-axis angle of
\cyg with respect to the instrument boresights of the AGILE-GRID (in red)
 and \fermi-LAT (in blue)   for the 10-day interval covering the period 18-28 July, 2009 with respect to  \s position for energies between 100 MeV and 400 MeV.
 During this interval the position of \cyg was stable near 30 degree off-axis in
 the AGILE data. The \fermi-LAT pointing strategy is very different from the AGILE fixed
 pointing mode strategy, and the \fermi-LAT off-axis angle changes continuously and rapidly
 for a typical 7-fold sampling of the source every day with a continuous sweep of
 the pointing direction. This difference in source pointing for \cyg in July 2009
 is clearly shown in the bottom panel of
 Fig. \ref{fig-11}, where the cumulative effective area of AGILE-GRID is greater than the \fermi-LAT (we have used the Pass 6 Version 3 Front photons Instrument Response Function).  The cumulative exposure of the two instruments operating with the same pointing strategy with respect to \s position is still comparable (AGILE-GRID effective area is about half that of the \fermi-LAT). Fig. \ref{fig-12} shows the case where AGILE operates in spinning mode. 
 
Both the AGILE-GRID average flux above 100 MeV and spectral index are in
good agreement with those reported in the Fermi-LAT First Catalog \cite{1FGL}
but difficult to reconcile with the results published in \cite{Abdo_2009b} and
\citep{corbel2010a} \abh{who report an average flux above 100 MeV of $F \simeq 50   \rm \, \times 10^{-8} ph. \, cm^{-2} \,  s^{-1}$ at the location of Cygnus X-3 outside the active $\gamma$-ray period, consistent with a stable emission}.
This relatively large flux would have produced a much
stronger stable $\gamma$-ray source in the AGILE-GRID data of the Cygnus region.

\section{Conclusions}
\abe{Our present study of the $\gamma$-ray emission from the microquasar \cyg adds substantially to
   the information already gathered from the first results of AGILE-GRID \citep{Tavani_2009b} and \fermi-LAT
  \citep{Abdo_2009b}.}
We confirm the time variable nature of the extreme particle
acceleration in the microquasar \cyg that manifests itself as $\gamma$-ray emission
above 100 MeV. 

The overall anticorrelation between the active $\gamma$-ray states
of \cyg and its hard X-ray emission is evident in our data that
confirm the conclusions of \citep{Tavani_2009b}. Both the AGILE pointing and spinning mode
observations of the Cygnus region sampled \cyg in different
spectral states, both ``hard'', and ``soft''. Enhanced $\gamma$-ray
emission is definitely produced only during the  ``soft''
spectral states. 
These states are
usually characterized by low values of the radio emission before
major radio outbursts occur.

We detected several $\gamma$-ray flaring episodes from \cyg with
fluxes above $ F = 100 \times 10^{-8} \, \phcmsec$ for photons
energies larger than 100 MeV. All these episodes occurred during
\abe{a low flux in hard X-rays}, and were simultaneous with either peak
values of the X-ray intensity or  low values of the radio flux
that typically precede major plasmoid ejections and radio flares.
Several episodes of enhanced $\gamma$-ray emission with
radio flares were detected  (MJD 54998, 55001, 55003, 55007,
55019, 55034, 55343 and possibly 55025).  In the majority of
cases, (except MJD 55323 that is seems correlated with a fast
transition between X-ray states) the  $\gamma$-ray emission above
100 MeV precedes or is near the major radio flare during  a
rapid spectral change transition.

The AGILE monitoring of \cyg in the spinning mode produced a
 consistent database  with no significant data gaps since 2009 Nov. We have continuously sampled
many X-ray state changes. \cyg stayed in the hard X-ray
state most of the time since 2009 Nov. (see Fig.
\ref{fig_1_2010}), implying a non-detectable $\gamma$-ray emission
from \cyg by AGILE. \abh{We finally stress that the AGILE monitoring without temporal gap is crucial for the comprehension of these transient phenomena}.

For the first time (around MJD 55000),
we witnessed the transition from high to low flux in hard X-ray,
with the consequent ``ignition'' of the $\gamma$-ray emission
process. Whether this is really an ``ignition'' , i.e., an enhanced and very
efficient particle acceleration process that start with the $\gamma$-ray production, or
a particle and photon transport  change of the medium surrounding the
\cyg compact source, is a matter of debate. AGILE will continue to study \s in order to discriminate between theoretical models.

\section{Acknowledgements}
The AGILE Mission is funded by the Italian Space Agency (ASI) with
scientific and programmatic participation by the Italian Institute
of Astrophysics (INAF) and the Italian Institute of Nuclear
Physics (INFN). Research partially supported through the ASI
grants no. I/089/06/2 and I/042/10/0. We acknowledge financial contribution from the agreement ASI-INAF I/009/10/0. The Mets\"ahovi team
acknowledges the support from the Academy of Finland for our
observing projects (numbers 212656, 210338, and others).


\newpage

\begin{thebibliography}{00}





\bibitem[ ]{ }
\bibitem[Abdo et al. 2009a]{Abdo_2009a}  Abdo, A.A., et al. (\fermip  Collaboration) 2009, Science, 325, 840
\bibitem[Abdo et al. 2009b]{Abdo_2009b}  Abdo, A.A., et al. (\fermip  Collaboration) 2009, Science, 326, 1512
\bibitem[Abdo et al. 2010]{1FGL}  Abdo, A.A. et al. (\fermip  Collaboration) 2010, ApJS 188, 405
\bibitem[Aleksic et al. 2010]{magic} Aleksic J. et al. (MAGIC Collaboration) 2010, ApJ, 721, 843-855
\bibitem[Argan et al. 2004]{argan} Argan, A., et al. 2004, Proc. IEE-NSS, 1, 371
\bibitem[Barbiellini et al. 2002]{Barbiellini_2002}  Barbiellini, G., et al. 2002, Nucl.~Instr.~and~Meth.~A, 490, 146
\bibitem[Benjamini et al. 1995]{Benjamini_1995} Benjamini, Y., Hochberg, Y. 1995, J. R. Stat. Soc. B, 57, 289
\bibitem[Bulgarelli et al. 2009]{Bulgarelli_2009} Bulgarelli, A., et al. 2008, ASP Conference Series, 411, 362
\bibitem[Bulgarelli et al. 2010]{Bulgarelli_2010} Bulgarelli, A., et al. 2010, Nucl.~Instr.~and~Meth.~A, 614, 213
\bibitem[Bulgarelli et al. submitted]{Bulgarelli_2011} Bulgarelli, A., et al., 2011 submitted.
\bibitem[Bulgarelli et al. 2010a]{atel2609} Bulgarelli, A., et al. 2010, ATel n. 2609
\bibitem[Bulgarelli et al. 2010b]{atel2645} Bulgarelli, A., et al. 2010, ATel n. 2645
\bibitem[Camilo et al. 2009]{camilo} Camilo, F., et al. 2009, ApJ, 705,1  (C09)
\bibitem[Cattaneo et al. 2011]{Cattaneo_2011} Cattaneo, P. W., et al. 2011, Proc. of the RICAP 2009, 251
\bibitem[Chen et al. 2011a]{2011A&A...525A..33C} Chen, A.~W., et al. 2011, A\&A,  525
\bibitem[Chen et al. 2011b]{Chen_2011b} Chen, A. W., et al. 2011b, in prep.
\bibitem[Chi et al. 1991]{Chi_1991} Chi X., Wolfendale A.W. 1991, J.Phys.G 17,987
\bibitem[Corbel et al. 2010a]{corbel2010a} Corbel, S., et al. 2010, ATel n. 2611
\bibitem[Corbel et al. 2010b]{atel2646} {Corbel}, S. and {Hays}, E. 2010, ATel n. 2646
\bibitem[Dame et al. 2001]{Dame_2001} Dame T.M. et al. 2001, ApJ, 547, 792
\bibitem[Edelson et al. 1988]{Edelson_1988} {Edelson}, R.~A., {Krolik},  J.~H. 1988, ApJ, 333, 646
\bibitem[Feroci et al. 2007]{Feroci2007} Feroci, M., et al. 2007, Nucl.~Instr.~and~Meth.~A, 581, 728
\bibitem[Giacconi et al. 1967]{giacconi} Giacconi, R., et al. 1967, ApJ, 148, L119
\bibitem[Grove et al. 2008]{grove} Grove, E., et al. 2008, ATel n. 1850
\bibitem[Halpern et al. 2008]{halpern} Halpern, J.P., et al. 2008, ApJ, 688, L33
\bibitem[Hjalmarsdotter et al. 2009]{hjalmarsdotter} Hjalmarsdotter, L., Zdziarski, A.A., Szostek, A., Hannikainen, D.C. 2009, MNRAS,  392, 251
\bibitem[Kalberla et al. 2005]{2005A&A...440..775K} {Kalberla}, P.~M.~W. , et al. 2005, \aa, 440:775--782
\bibitem[Koljonen et al. 2010]{koljonen} Koljonen, K.I.I., Hannikainen, D.C., McCollough, M.L., Pooley, G., Trushkin, S.A. 2010, MNRAS, 406, 307
\bibitem[Labanti et al. 2006]{Labanti_2006} Labanti, C., et al. 2006, Nuclear Physics B Proc. Suppl., 150, 34
\bibitem[Giuliani et al. 2004]{Giuliani_2004} Giuliani, A., et al. 2004, Memorie della Societa Astronomica Italiana Supplementi, 5, 135
\bibitem[Giuliani et al. 2006]{Giuliani_2006} Giuliani, A., et al. 2006, Nucl.~Instr.~and~Meth.~A, 568, 692
\bibitem[Hopkins et al. 2002]{Hopkins_2002} Hopkins, A.M., et al. 2002, AJ, 123, 1086
\bibitem[Mason et al. 1986]{mason} Mason, K.O., Cordova, F.A., and White, N.E. 1986, ApJ, 309, 700
\bibitem[Mattox et al. 1996]{Mattox_1996a} Mattox, J. R., et al. 1996, ApJ, 461, 396
\bibitem[Mattox et al. 2001]{Mattox_2001} Mattox, J. R., Hartman, R. C., Reimer, O. 2001, ApJ, 135, 155
\bibitem[McCollough et al. 1999]{mccollough} McCollough, M.L., et al. 1999, ApJ, 517, 951
\bibitem[McLaughlin et al. 1996]{0004-637X-473-2-763} McLaughlin, M.~A. , et al. 1996,  ApJ, 473(2):763
\bibitem[Miller et al. 2001]{Miller_2001} Miller, C. J. et al. 2001, ApJ, 122, 349
\bibitem[Mioduszewski et al. 2001]{miodu} Mioduszewski A.J., Rupen, M.P., Hjellming, R.M., Pooley, G.G. 2001, ApJ, 553, 766
\bibitem[Molnar et al. 1988]{molnar} Molnar, L.A., Reid, M.J., and Grindlay, J.E. 1988, ApJ, 331, 494
\bibitem[Parsignault et al. 1972]{parsignault} Parsignault, D.R., Gurski, H., Kellog, E.M., et al. 1972, Nature, 239, 123
\bibitem[Perotti et al. 2006]{Perotti_2006} Perotti, F., et al. 2006, Nucl.~Instr.~and~Meth.~A, 556, 228
\bibitem[Piano et al. in prep]{Piano_2011} Piano, G., et al., in prep.
\bibitem[Pittori et al. 2009]{Pittori_2009} Pittori, C., et al. 2009, \aa, 506, 1563.
\bibitem[Prest et al. 2003]{Prest_2003} Prest, M., et al. 2003,  Nucl.~Instr.~and~Meth.~A, 501,  280.
\bibitem[Sabatini et al. 2010]{Sabatini_2010} Sabatini, S., et al. 2010, \apjl, 712, L10--L15.
\bibitem[Sanford et al. 1972]{Sanford_1972} Sanford, P.~W.  and Hawkins, F.~H. 1972, Nature Phys Sci, 239, 135
\bibitem[Szostek et al. 2008]{szostek} Szostek, A., Zdziarski, A.A., McCollough, M.L. 2008, MNRAS, 388, 1001
\bibitem[Tavani et al. 2009a]{Tavani_2009a} Tavani, M.,  et al. 2009a, \aa, 520, 995
\bibitem[Tavani et al. 2009b]{Tavani_2009b} Tavani, M., et al. 2009b, Nature, 462, 620
\bibitem[van Kerkwijk et al. 1992]{vankerkwijk} van Kerkwijk, M.H., Charles, P.A., Geballe, T.R., et al. 1992, Nature, 355, 703
\bibitem[van Kerkwijk et al. 1996]{vankerkwijk2} van Kerkwijk, M.H.,  Geballe, T.R., King, D.L., van der Klis, M., van Paradijs, J. 1996, \aa, 314, 521
\bibitem[Vilhu et al. 2009]{vilhu} Vilhu, O. Hakala, P., Hannikainen, D., McCollough, M., \& Koljonen, K. 2009, A\&A, 501, 679
\bibitem[Williams et al. 2011]{Williams_2011} Williams, P.~K.~G., et al. 2011, \apjl, 733:L20
\bibitem[Zdziarski \& Gierlinski 2004]{zdziarski} Zdziarski, A.A. \& Gierlinski, M. 2004, Prog. Theor. Phys. Suppl., 155, 99

\end{thebibliography}
\end{document}